\documentclass[fleqn,usenatbib]{mnras}
\usepackage{type1cm} % Esto hace que podamos usar los tamaños impuestos por mnras sin problemas.
\usepackage{newtxtext,newtxmath}
\usepackage[utf8]{inputenc}

\usepackage[T1]{fontenc}
\usepackage{ae,aecompl}

\usepackage{graphicx}	% Including figure files
\usepackage{amsmath}	% Advanced maths commands
\usepackage{amssymb}	% Extra maths symbols
    
\usepackage[flushleft]{threeparttable}

\title[LAEs and LBGs in SHARDS]{A simultaneous search for High-$z$ LAEs and LBGs in the SHARDS survey}

\author[P. Arrabal Haro et al.]{P. Arrabal Haro$^{1,2}$\thanks{E-mail: parrabal@iac.es}, J. M. Rodríguez Espinosa$^{1,2}$, C. Muñoz-Tuñón$^{1,2}$,
\newauthor P. G. Pérez-González$^{3}$, H. Dannerbauer$^{1,2}$, Á. Bongiovanni$^{1,2}$, G. Barro$^{4}$, A. Cava$^{5}$,
\newauthor A. Lumbreras-Calle$^{1, 2}$, A. Hernán-Caballero$^{3}$, M. C. Eliche-Moral$^{1}$,
\newauthor H. Domínguez Sánchez$^{6}$, C. J. Conselice$^{7}$, L. Tresse$^{8}$, B. Alcalde Pampliega$^{3}$,
\newauthor M. Balcells$^{9,1,2}$, E. Daddi$^{10}$ and G. Rodighiero$^{11}$
\\
$^{1}$Instituto de Astrofísica de Canarias (IAC), E-38205 La Laguna, Spain\\
$^{2}$Departamento de Astrofísica, Universidad de La Laguna, E-38206 La Laguna, Spain\\
$^{3}$Departamento de Astrofísica y CC de la Atmósfera, Universidad Complutense de Madrid, E-28040 Madrid, Spain\\
$^{4}$Department of Physics, University of the Pacific, 3601 Pacific Avenue Stockton, CA 95211, USA\\
$^{5}$Department of astronomy, University of Geneva, 51 Ch. des Maillettes,
CH-1290 Versoix, Switzerland\\
$^{6}$Department of Physics and Astronomy, University of Pennsylvania,
Philadelphia, PA 19104, USA\\
$^{7}$School of Physics \& Astronomy, University of Nottingham, Nottingham NG7 2RD, UK\\
$^{8}$Univ Lyon, Univ Lyon1, Ens de Lyon, CNRS, Centre de Recherche Astrophysique de Lyon (CRAL) UMR5574, F-69230 Saint-Genis-Laval,\\ France\\
$^{9}$Isaac Newton Group of Telescopes, Apartado 321, 38700 Santa Cruz de La Palma, Spain\\
$^{10}$Laboratoire AIM-Paris-Saclay, CEA/DSM-CNRS-Université Paris Diderot, Irfu/Service d'Astrophysique, CEA Saclay, Orme des Merisiers,\\ F-91191 Gif-sur-Yvette, France\\
$^{11}$Dipartimento di Fisica e Astronomia, Università di Padova, Vicolo dell’Osservatorio 3, I-35122, Italy}

\date{Accepted 2018 April 24. Received 2018 April 3; in original form 2018 January 20.}

\pubyear{2018}

\begin{document}
\label{firstpage}
\pagerange{\pageref{firstpage}--\pageref{lastpage}}
\maketitle

% Abstract of the paper. It should be a single paragraph not more than 250 words (200 words for Letters). No references should appear in the abstract.
\begin{abstract}
We have undertaken a comprehensive search for both Lyman Alpha Emitters (LAEs) and Lyman Break Galaxies (LBGs) in the SHARDS Survey of the GOODS-N field. SHARDS is a deep imaging survey, made with the 10.4 m Gran Telescopio Canarias (GTC), employing 25 medium band filters in the range from 500 to 941 nm. This is the first time that both LAEs and LBGs are surveyed simultaneously in a systematic way in a large field. We draw a sample of 1558 sources; 528 of them are LAEs. Most of the sources (1434) show rest-frame UV continua. A minority of them (124) are pure LAEs with virtually no continuum detected in SHARDS. We study these sources from $z\sim3.35$ up to $z\sim6.8$, well into the epoch of reionization. Note that surveys done with just one or two narrow band filters lack the possibility to spot the rest-frame UV continuum present in most of our LAEs. We derive redshifts, Star Formation Rates (SFRs), Ly$\alpha$ Equivalent Widths (EWs) and Luminosity Functions (LFs). Grouping within our sample is also studied, finding 92 pairs or small groups of galaxies at the same redshift separated by less than 60 comoving kpc. In addition, we relate 87 and 55 UV-selected objects with two known overdensities at $z=4.05$ and $z=5.198$, respectively. Finally, we show that surveys made with broad band filters are prone to introduce many unwanted sources ($\sim20$\% interlopers), which means that previous studies may be overestimating the calculated LFs, specially at the faint end.
\end{abstract}

%%%%%%%%%%%%%%%%%%%%%%%%%%%%%%%%%%%%%%%%%%%%%%%%%%

%%%%%%%%%%%%%%%%% BODY OF PAPER %%%%%%%%%%%%%%%%%%

\section{Introduction}
\label{sec:Introduction}
The reionization epoch is an important phase in the evolution of the universe that is still not completely understood. The currently favoured model is that the transition from the neutral phase to a fully ionized universe was achieved by an abundant population of low luminosity star-forming galaxies \citep{Ouchi2009,Bouwens2011} from $z\sim15$ to about $z\sim5.5$ \citep{Fan2006,Robertson2010}. 
These sources are the best tracers of star formation at high redshifts and are commonly separated into two groups depending on whether or not they show Ly$\alpha$ in emission. The so called  Lyman Alpha Emitters (LAEs) show a prominent Ly$\alpha$ emission line, while those showing strong rest-frame UV continuum and a Lyman break are called Lyman Break Galaxies (LBG). The latter may or may not show Ly$\alpha$ in emission. In this paper we will use LAEs for any galaxy with a Ly$\alpha$ emission line, while LBGs will be used for any galaxy with the Lyman break and a well defined UV continuum at longer wavelengths. Traditionally, LAEs have been detected using narrow band filters while broad band filters are used to detect LBGs. In addition, multi-object or integral field spectroscopy have been used recently to successfully detect high-$z$ sources \citep[\textit{e.g.},][]{Drake2017, Herenz2017}. Many previous studies have dealt with both types of objects at different redshifts \citep[\textit{e.g.,}][among others]{Hu1998,Shapley2003,Steidel2003,Giavalisco2004,Malhotra2004,Steidel2005,Fan2006,Gronwall2007,Iwata2007,Ouchi2008,McLure2009,Oesch2010,Ouchi2010,vdB2010,Bouwens2014,Bouwens2015,Cassata2015,Sobral2017,Sobral2017b}. However, there are very few spectroscopic data as these sources are quite faint, thus only the brightest can be detected \citep[\textit{e.g.},][]{Caruana2014, Caruana2018}. In addition, because of their faintness, it is also difficult to obtain good spectral energy distributions (SEDs) and thus to remove low redshift interlopers.\\

To overcome these problems we use the Survey for High-$z$ Absorption Red and Dead Sources (SHARDS), described in detail in \citet{PabloSHARDS}. This survey studies the Great Observatories Origins Deep Survey North (GOODS-N) field, employing 25 consecutive medium width filters, between 500-941 nm. The major advantage of the work presented here over previous ones is that SHARDS allows the simultaneous detection of both LAEs and LBGs in a uniform way, as \citet{Bina2016}, \citet{Drake2017} and \citet{Drake2017b} did on smaller fields using MUSE (Multi Unit Spectroscopic Explorer). Furthermore, the 25 medium-band filters provide much more accurate SEDs than broad band filters, facilitating the rejection of interlopers. The wavelength range sampled makes this survey ideal to follow the evolution of LAEs and LBGs from $z\sim3.35$ to $z\sim6.8$. Even though SHARDS does reach quite faint depths (26.5-27.0 AB mag), it does not reach the depth of broad band studies carried out with the Hubble Space Telescope (HST), which can go up to $\sim29.0$-29.5 AB mag in some cases. Nonetheless, for those galaxies within our magnitude limit, we do have much better information about their observed optical/NIR SEDs, thus we can perform a more reliable physical characterization.\\

Another interesting aspect of the sources responsible for the reionization is their grouping \citep{McQuinn2007}. Recently, many authors have found proto-clusters at high redshifts \citep[\textit{e.g.,}][]{Steidel1998, Venemans2002, Venemans2007, Blain2004,Daddi2009,Mancini2009,Capak2011,Toshikawa2012,Toshikawa2014,Walter2012,Cucciati2014,Helmut2014,Lemaux2017}. These may set important constraints to model large-scale structure formation. Moreover, clustering is more important at higher redshifts, where the neutral hydrogen is still very abundant in the universe. Therefore, in order to detect Ly$\alpha$ emission, a big ionized gas bubble is needed around those galaxies. Many ionizing galaxies in the same region contribute to create those bubbles, which eventually will grow and merge to fully complete the reionization of the universe.\\

This work is structured as follows: in Section~\ref{sec:Database} we briefly describe the SHARDS survey and the Rainbow Cosmological Surveys Database and explain the process used to carry out our candidates selection, as well as to eliminate interlopers; in Section~\ref{sec:Physical_parameters} we calculate several physical parameters; in Section~\ref{sec:Discussion} we present the relevant results and compare them with previous studies; finally, in Section~\ref{sec:Conclusions} we summarize the main conclusions. In what follows all calculations were made adopting a $\Lambda$-dominated flat universe with $H_{0}=68\ \mathrm{km\ s^{-1}\ Mpc^{-1}}$, $\Omega_{\mathrm{M}}=0.3$ and $\Omega_{\mathrm{\Lambda}}=0.7$ \citep{Planck2016}. All magnitudes are expressed in the AB system and all physical distances refer to comoving distances.

\section{The data}
\label{sec:Database}
\subsection{The SHARDS survey and the Rainbow Cosmological Surveys Database}
\label{sec:SHARDS_description}
SHARDS is an ESO/GTC deep optical spectro-photometric survey of the GOODS-N field acquired with 200 hours of observing time with the OSIRIS instrument on the 10.4 m Gran Telescopio Canarias (GTC). The surveyed area is $\sim130$ arcmin$^{2}$, split into two pointings. This survey provides very deep photometry ($m\lesssim26.5-27.0$ AB mag, at the $3\sigma$ level) in 25 medium-band filters, from 500 nm to 941 nm. The first 22 filters have $\sim17$ nm FWHM, while the last three have a FWHM of 35, 25 and 33 nm, respectively.\\

The main purpose of SHARDS was to provide SEDs with good enough spectral resolution to study red and passive galaxies up to redshift $z\sim2$. Nonetheless, they can be used to detect line emitting sources that otherwise would not be detected using broader filters \citep{Cava2013,J&C2014,Hernan2017,Lumbreras2018}, as well as drop-out sources that can be well characterized by the many consecutive filters. Indeed, the relative narrowness of the SHARDS filters and large number of them result in good spectrally resolved SEDs ($R\sim50$). This is what makes SHARDS an excellent survey to search and study LAEs and LBGs at redshifts between 3.35 and 6.8, whose Ly$\alpha$ emission line or Lyman break fall within the wavelength range of SHARDS. The SHARDS data reduction and calibration is explained in great detail in \citet{PabloSHARDS}.\\

We have also made use of additional public data from \textit{HST}/ACS \citep{Giavalisco2004b,Riess2007}, \textit{HST}/WFC3 \citep{Grogin2011,Koekemoer2011}, \textit{Subaru}/Suprime-Cam \citep[as a part of the Hawaii Hubble Deep Field North project,][]{Capak2004}, \textit{Subaru}/MOIRCS \citep{Kajisawa2011}, \textit{CFHT}/WIRCam \citep{Lin2012}, \textit{Spitzer}/IRAC \citep{Fazio2004,P-G2005,P-G2008,Ashby2015} and GALEX \citep{Bianchi2014} as compiled in the Rainbow Cosmological Surveys Database\footnote{Operated by the Universidad Complutense de Madrid (UCM), partnered with the University of California Observatories at Santa Cruz (UCO/Lick, UCSC).\\ \url{http://rainbowx.fis.ucm.es/Rainbow_navigator_public/}} \citep{Barro2011,Barro2011b,Barro2018}. We have therefore added valuable data in wavelengths beyond the SHARDS range. This is important to ensure that our candidates do not present significant emission bluewards of the Lyman break, and in general to discard interlopers.

\subsection{Data processing}
\label{sec:Data_proc}
We start by limiting our search field to the common area of all the SHARDS images, for each pointing, as each SHARDS image samples a slightly displaced area. Thus the detection code only takes into account sources that are inside a resulting effective area of 128.4 arcmin$^{2}$ (otherwise we get many spurious detections near the borders of the frames).\\

We also need to take into consideration the effective central wavelength (CWL) within each filter, which depends on the particular position of each source in the field of view. This is a pure geometrical effect, as a result of the angle of incidence of the GTC/OSIRIS light beam on the filter. Objects in the same SHARDS image are detected with similar effective wavelengths but not exactly the same, depending on their precise position in the image. Fortunately, this is an effect already calibrated \citep[see][]{PabloSHARDS}, so we can get the exact CWL for each object in each filter directly from the SHARDS catalogue. The appropriate FWHM of the band sampling that CWL was also considered, as this varies slightly for each filter.

\subsection{Sample selection}
\label{sec:Sample_selection}
Here we explain in detail the process used to search for the candidate sources in the SHARDS catalogue. SHARDS contains 44,752  sources in the GOODS-N field, and for each object there is information on its flux in the SHARDS filters, plus the ancillary data available in Rainbow (see Section~\ref{sec:SHARDS_description}).\\

A first selection was made taking advantage of the photometric redshifts available in the Rainbow Database for 35,445 of the SHARDS objects in the CANDELS/GOODS-N field. These redshifts were obtained by \citet{Barro2018}, hereinafter Ba18, through SED fitting with EAZY \citep{Brammer2008}, using SHARDS combined with the WFC3 102 and 141 grisms and additional data from HST, Subaru, Spitzer and others. This results in a very well detailed SED characterization and redshift calculation, reaching an accuracy better than $\Delta z\sim0.01$ \citep[see also][]{Ferreras2014,Dominguez2016}. From this sample, 1078 candidates were selected using their redshift estimation.\\

\begin{figure}
	\includegraphics[width=\columnwidth]{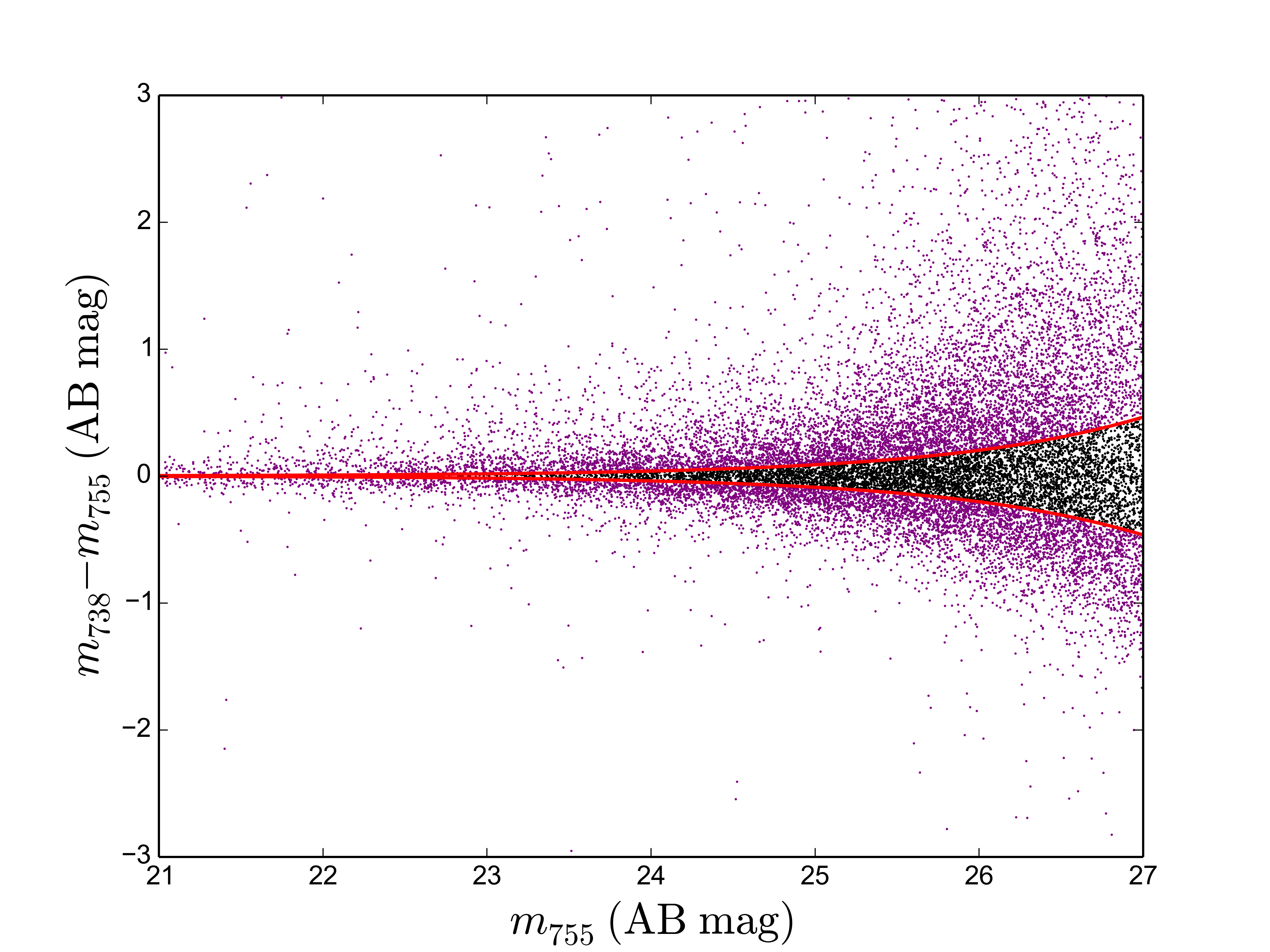}
    \caption{Example of a colour-magnitude diagram for the filters f738w17 and f755w17. The red lines represent $1\sigma$ errors for each magnitude, so they limit the objects we can reliably consider having an emission line in these filters. Purple points above the upper red line have higher magnitude in the f738w17 filter and therefore they show a reliable emission drop with respect to the f755w17.}
    \label{fig:C-M_diagram}
\end{figure}

For the 9,307 objects in the SHARDS catalogue, without  photometric redshift in Ba18, a selection criterion based on colour excesses between filters was designed. Since we are interested in detecting both Ly$\alpha$ in emission and/or the Lyman break feature, we  first constructed colour-magnitude diagrams using consecutive filters. Thus we were able to detect all relevant flux changes from one filter to the next (regardless of whether they were emission lines or drop-outs). In order to only select objects with reliable flux changes, we studied the dependence of  magnitude errors with magnitude in the SHARDS catalogue for each one of the filters. We then estimated the mean error for a given magnitude and determined whether a flux change between filters was large enough. If, on the contrary, this error was within the expected errors for that magnitude, we discarded the source. The procedure used first order logarithmic fits of the magnitude error for each filter. We used $1\sigma$ error limits to identify relevant emission changes (see Fig.~\ref{fig:C-M_diagram}).\\

We then built SHARDS SEDs for the objects that satisfied the colour-magnitude criterion, and started a more selective filtering. For this purpose we implemented a code that analyses the SEDs of the objects and selects candidates according to the following criteria:

\begin{equation}
\label{eq:selection_criteria}
 \qquad \qquad \delta\leq \langle m_{j<i}\rangle-m_{i} \wedge \Delta m_{i}<\langle m_{j<i}\rangle-m_{i},
\end{equation}\

\noindent where $m_{i}$ is the AB apparent magnitude of the emission filter just after the break, $\Delta m_{i}$, its error, $m_{j<i}$ are the magnitudes of all the bluer filters before $m_{i}$. Finally, $\delta$ is an adjustable magnitude difference for the break, which we kept above the $1\sigma$ level of the colour-magnitude diagram limits. In other words, our code looks for a source in filter $i$ whose emission is at least $\delta$ AB magnitudes brighter than its mean magnitude in all the non-zero emission filters bluewards of the said filter $i$ (first term of eq.~[\ref{eq:selection_criteria}]). The code keeps only those cases where the emission gap is larger than the errors in the magnitude of the brightest filter defining the break (second term of eq.~[\ref{eq:selection_criteria}]). 
We compared each filter detection with the mean detection of all the filters bluer than the original filter to make sure that what we detect is a spectral break, and not other absorption features affecting only one of the SHARDS bands. This method still produced a large number of high-$z$ candidate SEDs (1642), but a number that we can visually check. As expected, many of the detections ($\sim70$\%) were either spurious objects, high slope red galaxies/stars or simple emission line galaxies, not LAEs or LBGs. Nonetheless, this turned out to be a good method to get a first cut of objects, while not imposing such a restrictive constrain as to discard valid candidates.\\

The last step was to visually inspect all the candidate objects, not only looking at their SEDs but also directly viewing the images. In this process we used images mainly from \textit{GTC}/SHARDS, \textit{HST}/ACS, and \textit{HST}/WFC3, though we also used other images available in the Rainbow Database from \textit{Subaru}/Suprime-Cam and \textit{Spitzer}/IRAC. The SEDs were further completed with useful data at other wavelengths available in Rainbow (see Section~\ref{sec:SHARDS_description}). Finally, we visually discarded those objects which were deemed either spurious or presented strong emission blueward of their supposed Lyman break wavelength. This check is necessary since lower $z$ objects with a high slope SED could match our SHARDS preselection criteria when their emission in one of the redder filters is substantially higher than the average emission in the filters blueward of that one.\\

The whole process was iterated many times, identifying the wavelength where the Lyman break appears for each object, paying special attention to galaxies with photometric redshift estimations from previous authors, making sure we did not skip possible candidates while not including unwanted objects. Using this last method we added 492 objects. These together with the 1078 selected objects from Ba18 mentioned before constitute our final selection sample, consisting of 1570 well characterized candidates, pending a final interlopers screening.

\subsection{Interloper rejection}
\label{sec:Interloper_rejection}
Once the selection was finished, we carried out some additional theoretical models fits to our SEDs to make a further test on those objects that had met our selection criteria but could still be lower $z$ interlopers. In the search for high-$z$ LBGs the main interlopers are lower $z$ galaxies where the D4000 break is wrongly identified as the Lyman break. Additionally, M, L, and T dwarf stars with a very steep slope can be mistakenly taken as drop outs. In these, it is possible to measure an abrupt flux change, specially with broad band filters. However, these last cases are in general easily recognizable when using many filters.  So we did not expect an important number of them among our interlopers. The fitting process was made with the code \textit{Le Phare} \citep{Arnouts1999,Ilbert2006}. We used different galaxy models \citep[\textit{e.g.,}][]{Coleman1980, Kinney1996, Bruzual2003}, assuming a \citet{Salpeter1955} Initial Mass Function (IMF) and a Calzetti extinction law \citep{Calzetti1994}. For the stars rejection, we used SEDs from \citet{Bohlin1995}, \citet{Kinney1996} and \citet{Chabrier2000}, with a special attention to  cold M, L and T stars from \citet{Pickles1998}, \citet{Burgasser2004} and others.\\

\begin{figure}
	\includegraphics[width=\columnwidth]{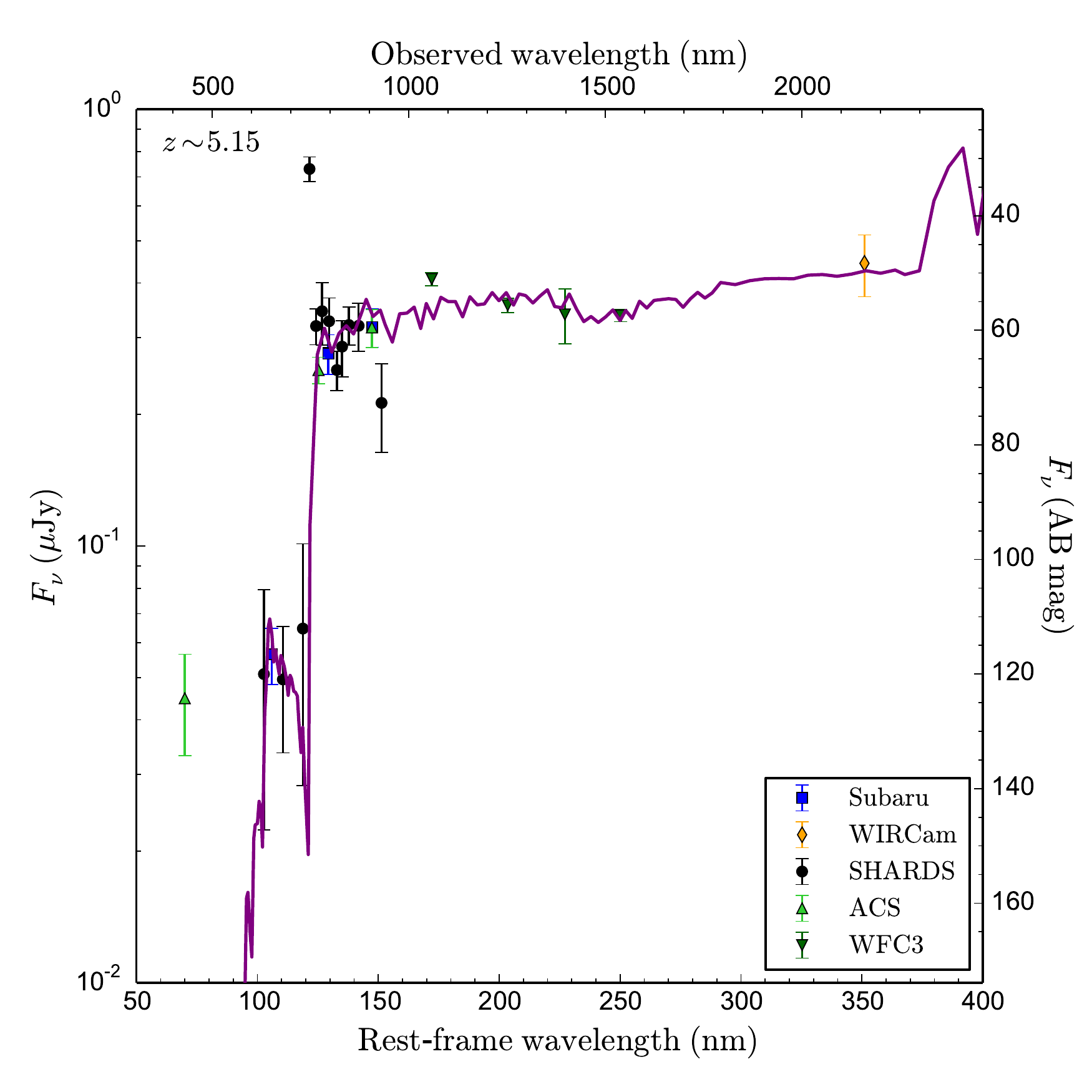}
    \caption{SED of the candidate SHARDS J123655.49+621532.7, a LBG/LAE with emission line at 747.5 nm corresponding to the Ly$\alpha$ line at redshift $z\sim5.15$. To make this SED we made use of all the data available in the Rainbow Database with reliable information on this galaxy. Notice that all emission blueward of the line goes down abruptly, while to the red the UV continuum can be easily seen. The purple line is a template of a $z\sim5.15$ model. All the SHARDS images of this object are shown in Fig.~\ref{fig:8819_image}.}
    \label{fig:8819_SED}
\end{figure}

\begin{figure*}
	\includegraphics[width=\textwidth]{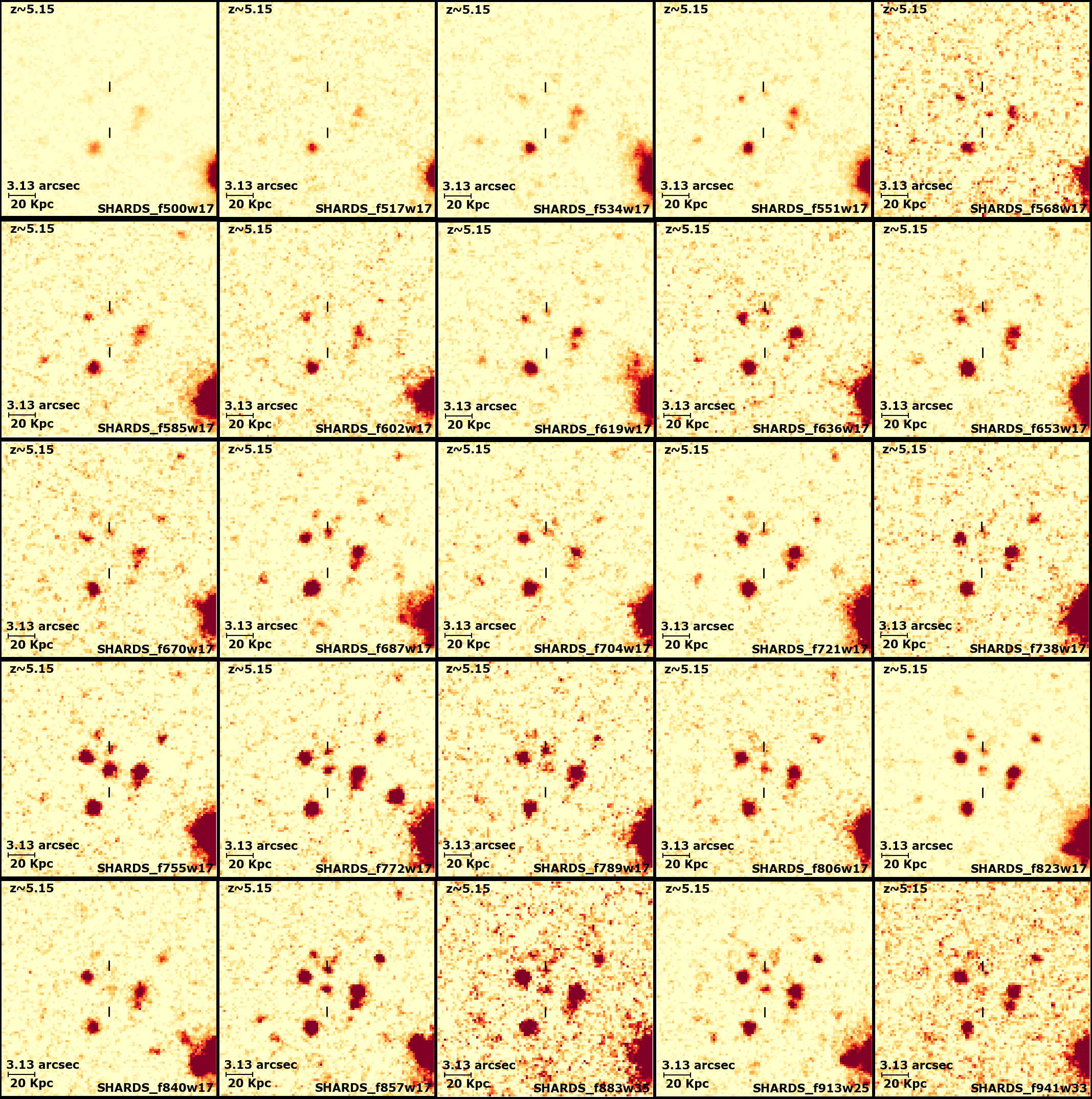}
    \caption{Mosaic made with the images of the 25 SHARDS filters from bluer (top left) to redder (bottom right) of the candidate source SHARDS J123655.49+621532.7 (enclosed between two marks). Its SED is shown in Fig.~\ref{fig:8819_SED}. The strongest emission, corresponding to the Ly$\alpha$ line, can be clearly appreciated in filter f755w17. Redder filters show fainter emission, corresponding to the UV rest-frame continuum, while shorter wavelengths filters do not show any emission at all at the object position. North is up, East is left.}
    \label{fig:8819_image}
\end{figure*}

As in every other survey, we did not get good SED fitting for all the 1570 candidate galaxies in our sample, but only for 644 (with $\chi^{2}\lesssim40$). Therefore, we  expected some unwanted objects within the worst fitted sources of the sample. Only 12 out of the well fitted part of the sample were found to be lower $z$ interlopers. None of them were found to be steep red-slope stars, but lower $z$ galaxies. To estimate the final interloper fraction in our high-$z$ galaxy selection we extrapolated the interloper fraction found for the well fitted sample to the whole sample, obtaining $\sim2$\% fraction of interlopers in our final selection. Thus, after eliminating the identified interlopers, our final sample consists of 1558 high-$z$ galaxies, classified as 528 LAEs and 1030 LBGs with no emission line. 124 of the emitters are pure LAEs with no continuum detected in SHARDS (although it can be measured in the \textit{HST}/ACS broad band images), while the rest (404) are LBGs with Ly$\alpha$ line emission. Figures~\ref{fig:8819_SED} and \ref{fig:8819_image} show examples of a good candidate SED and its view on each SHARDS filter, respectively.

\subsection{Completeness of the sample}
\label{sec:Completitud}
\begin{figure}
	\includegraphics[width=\columnwidth]{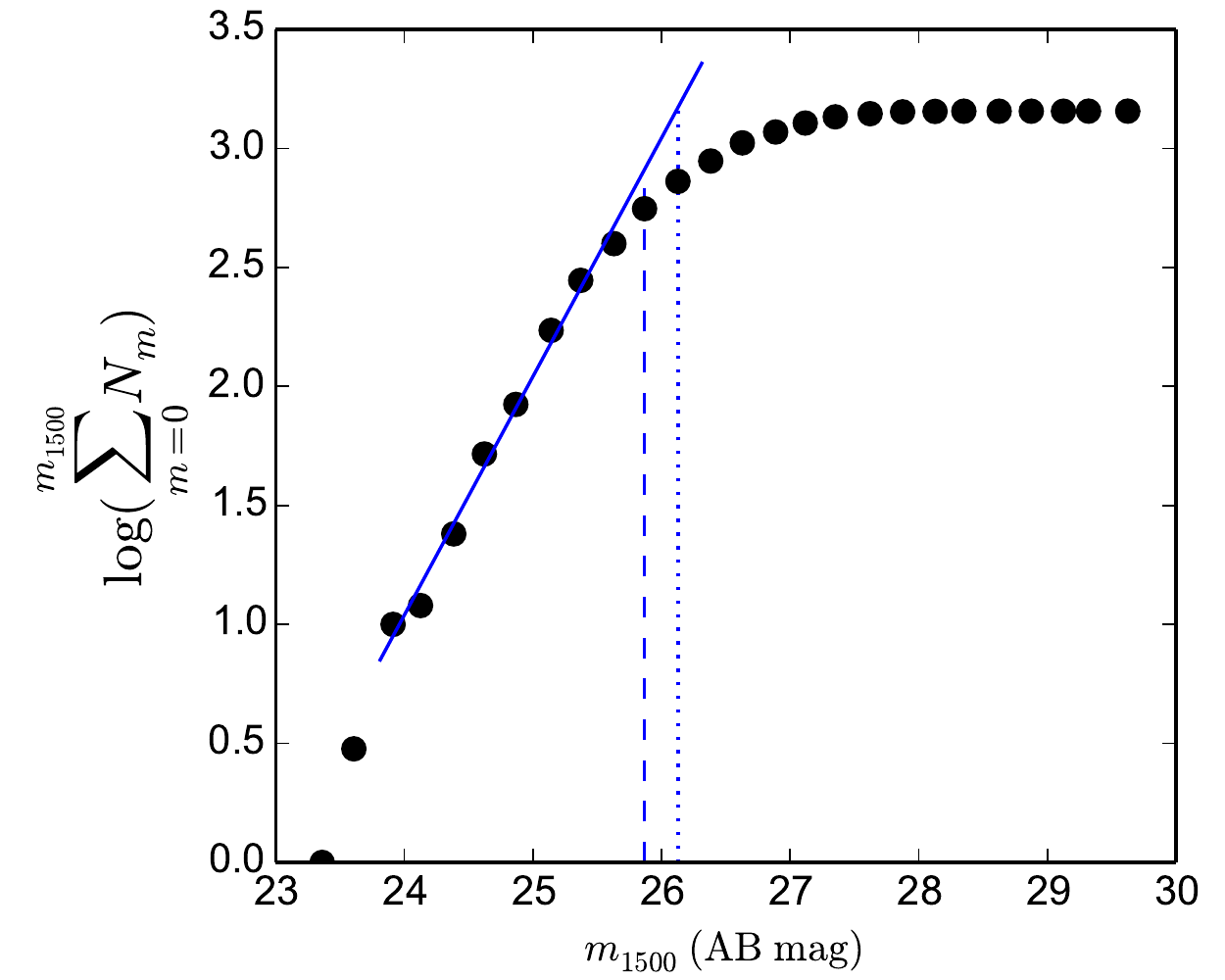}
    \caption{Logarithmic cumulative function of AB apparent magnitude of the LBG candidates. The point where the slope changes indicates the magnitude at which we start missing objects. The blue solid line is a fit to the slope while the dashed and dotted lines indicate the magnitudes at which we are 90\% and 50\% complete according to that fit, which in our case are $\sim25.87$ and $\sim26.13$ AB mag, respectively.}
    \label{fig:Completitud}
\end{figure}

In order to properly correct the Luminosity Functions (LF) and to compare our sample with previous ones, we estimated our completeness. To achieve this, we took the rest-frame UV magnitude at 1500 \AA{} as reference magnitude and made a plot of the logarithm of the cumulative number of LBGs with $m_{1500}$ below some apparent magnitude. As we do not reach faint enough magnitudes to sample the potential part of the LF, our selection will be mostly dominated by the exponential term of the Schechter function. It is thus reasonable to assume that we can calculate our completeness looking at the $m_{1500}$ at which the slope of our logarithmic cumulative distribution starts decreasing. In this way we obtain the value beyond which we start missing candidate objects. This will be our completeness apparent magnitude (see Fig.~\ref{fig:Completitud}). We obtain a 90\% completeness at magnitude $\sim25.87$ AB mag, although our faintest sources reach up to $m_{1500}\sim28$. This completeness value corresponds to a different absolute magnitude for each different redshift.

\section{Physical parameters}
\label{sec:Physical_parameters}
In this section we discuss the redshift distribution, Star Formation Rates (SFRs), Ly$\alpha$ Equivalent Widths (EWs), and grouping properties of our sample.

\subsection{Redshifts}
\label{sec:Redshift}
Section~\ref{sec:Sample_selection} showed how the SED of each object was used to identify the filter wavelength at which that object presents either an emission line or a continuum drop. The main assumptions, for obtaining the redshift of the objects, is that the line is Ly$\alpha$ and the emission drop is the Lyman break. These premises are supported by the selection criteria described in Sections~\ref{sec:Sample_selection} and \ref{sec:Interloper_rejection}. With these assumptions, we get their break wavelength, taking into account the precise position in the image to correct the CWL shift effect described in Section~\ref{sec:Data_proc}. Our preliminary calculations pointed out that this effect is relevant when the object is located far from the central region. We also take care of using the appropriate filter width corresponding to the pre-break filter. It is important to clarify that the Lyman alpha break wavelength used is not 912 \AA{} but just the blue side of Ly$\alpha$, since the Ly$\alpha$ forest due to intergalactic absorption erases almost any continuum emission blueward of 1215.7 \AA{} at high redshifts \citep{Rauch1998,Fan2006b}.\\

\begin{figure}
	\includegraphics[width=\columnwidth]{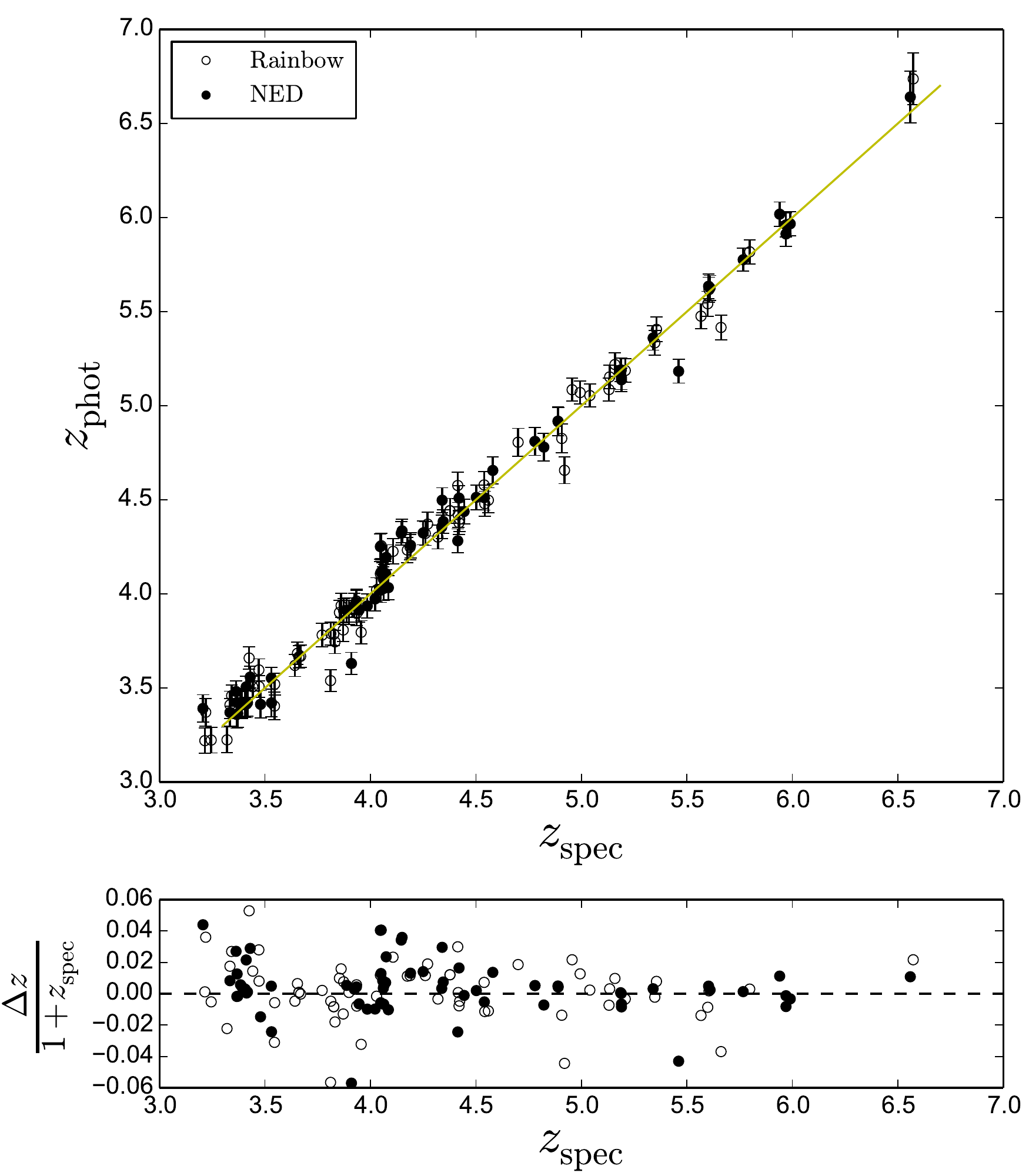}
    \caption{Comparison between our photometric redshifts with the spectroscopic ones for those candidates with spectroscopic $z$ available in Rainbow or NED. Black dots are extracted from NED while empty circles are from Rainbow. In case the same object has spectroscopic $z$ in both databases we plot the NED one only to avoid redundancy (after checking that the value is the same). The solid line is the equal line. Notice that none of the spectroscopically confirmed objects of our sample has a calculated photo-$z$ over the outliers threshold ($|\Delta z|/(1+z_\mathrm{spec})=0.15$). The excellent correlation indicates that both our selection criteria and redshift calculations are very robust and accurate.}
    \label{fig:Zspec_Zphot}
\end{figure}

To prove the goodness of our photometric redshifts we check every case where a spectroscopic redshift was available either in the Rainbow Database or in the NASA/IPAC Extragalactic Database (NED)\footnote{\url{https://ned.ipac.caltech.edu/}} \citep[from][among others]{Steidel2003,Kakazu2007,Barger2008,Kajino2009,Adams2011,Conselice2011,Shim2011,Stark2011,Stark2013,Schenker2013} and compare them with our photometric redshifts (see Fig.~\ref{fig:Zspec_Zphot}). The excellent correlation obtained indicates that we can trust the calculated redshifts for our candidates with an average error of around $\Delta z=0.07$ (0.14 for objects above $z\sim6.2$, where the filters measuring the Lyman break are wider).\\

\begin{figure}
	\includegraphics[width=\columnwidth]{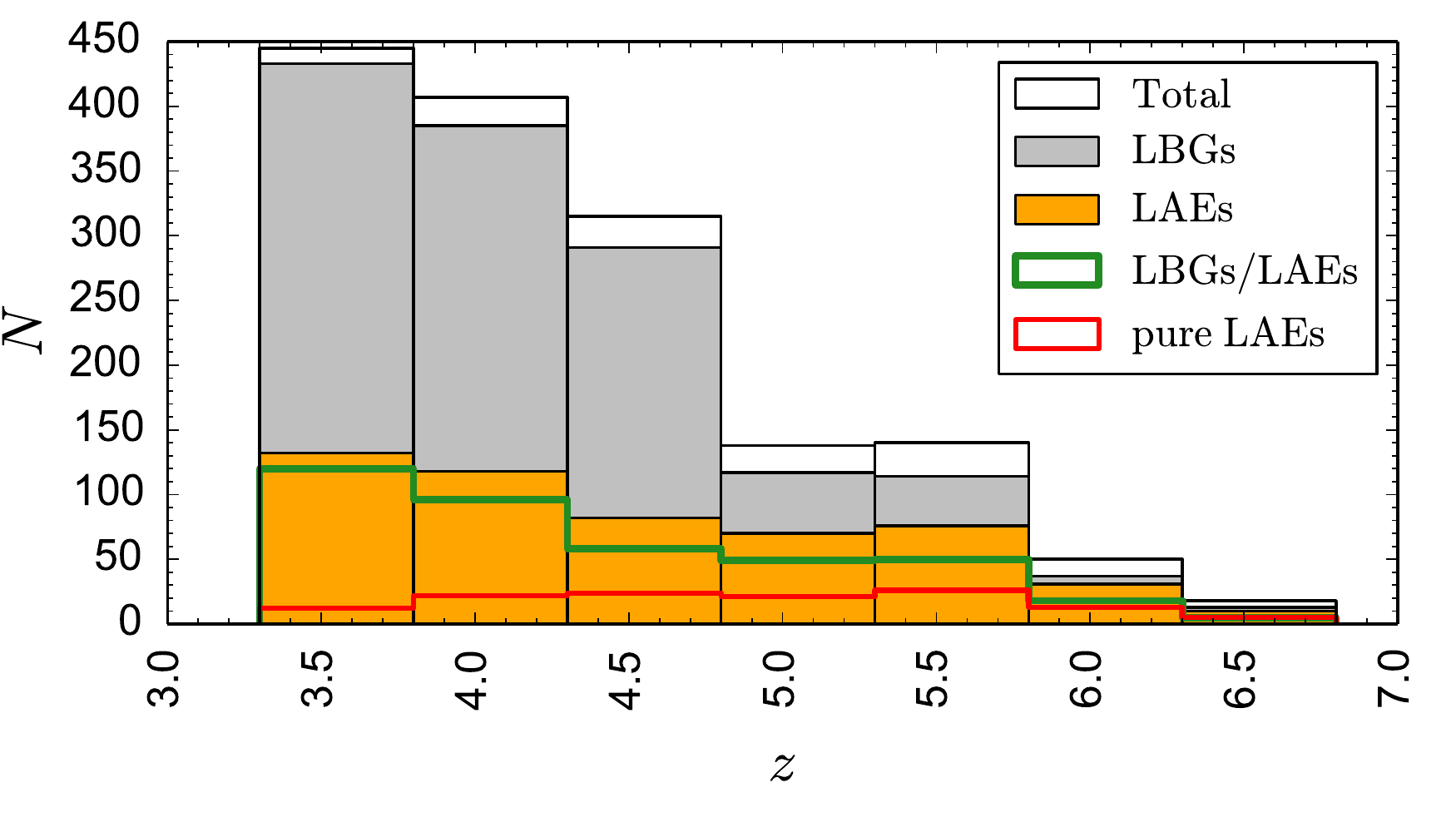}
    \caption{Redshift distribution of the sample. In grey we plot the LBGs. In orange, the LAEs. These last ones are divided in LBGs/LAEs with UV continuum detection in SHARDS, represented in green, and pure LAEs with no continuum detection in SHARDS, in red.}
    \label{fig:Hist_Z}
\end{figure}

The redshift distribution of our LAE and LBG candidates is shown in Fig.~\ref{fig:Hist_Z}. The number of LBGs decreases with $z$, which may be partly due to the lower number of SHARDS filters providing relevant information in the red part of the spectrum, apart from the Malmquist bias, as farther away objects are fainter. We also notice that the relative number of LAEs is larger beyond $z\sim5.0$-5.5. The fraction of pure LAEs and LBGs and their behaviour are discussed in depth in \citet{Arrabal2018letter}. The complete list of redshifts is in the on-line version.\\

Two previously reported GOODS-N overdensities at $z\sim4.05$ \citet{Daddi2009} and $z\sim5.2$ \citet{Walter2012} have also been recovered in this work. The number of objects within our sample which may belong to those overdensities according to their redshifts are 87 and 55, respectively. For the $z\sim5.2$ proto-cluster, those authors find 13 objects at $z=5.198\pm0.015$ in the GOODS-N field, 11 of which are also detected by us. Of the two not detected, one is too faint and the other one, HDF850.1, was detected via cold molecular gas lines \citep{Walter2012}, not showing any emission in the optical/NIR. We have further detected 44 additional objects whose redshifts are compatible with $z=5.198\pm0.015$, within the errors. They are all detected with the Ly$\alpha$ line/break in the SHARDS filter f755w17 and have an estimated redshift error of 0.07. Although we have to wait for spectroscopy of these sources to further constrain their redshifts, a total of 55 possible cluster members for a single proto-cluster is something unseen beyond z=5 \citep[\textit{cf.}, \textit{e.g.},][]{Capak2011,Toshikawa2012,Toshikawa2014,Higuchi2018}, making this a very interesting overdensity.

\subsection{Star Formation Rates}
\label{sec:SFR}
Traditionally, two different methods have been used to obtain SFRs for the LAEs and LBGs respectively. The emitters SFR can be obtained from the Ly$\alpha$ line emission following \citet{Kennicutt1998}, who assumed case B recombination \citep{Brocklehurst} and a Salpeter IMF \citep{Salpeter1955}, through the relation:

\begin{equation}
 \qquad \qquad \mathrm{SFR} \ (M_{\odot} \ \mathrm{yr^{-1}})=\frac{L(\mathrm{H}\alpha) \ (\mathrm{erg \ s^{-1}})}{1.26\times10^{41}},
\end{equation}
\\
where $L(\mathrm{H}\alpha)=L(\mathrm{Ly}\alpha)/8.7$.\\

The other method is based on the UV continuum luminosity density at 1500 \AA{} ($L_{1500}$) with the assumptions of solar metallicity and a Salpeter IMF, as in \citet{Madau1998}:

\begin{equation}
\label{eq:Madau}
 \qquad \mathrm{SFR} \ (M_{\odot} \ \mathrm{yr^{-1}})=1.25\times10^{-28}L_{1500} \ (\mathrm{erg \ s^{-1} \ Hz^{-1}}).
\end{equation}
\\
To correct for galactic dust extinction, $A_{\lambda}$ values from the \citet{Schlafly2011} dust maps are used. The internal dust extinction is calculated following \citet{Calzetti1994}, \citet{Calzetti2000} and \citet{Castellano2012}, assuming $F_{\mathrm{\lambda}}\propto\lambda^{\beta}$, where $\beta$ is the UV slope. This $\beta$ slope is estimated via linear fit to the magnitudes measured in the filters sampling the rest-frame UV wavelength from 1300 \AA{} through to 2600 \AA{}:

\begin{equation}
 \qquad\qquad\qquad m_{i}=-2.5(\beta+2.0)\log(\lambda_{i})+C,
\end{equation}
\\
\noindent where $m_{i}$ is the magnitude measured in the filter centred in $\lambda_{i}$ and $C$ is a constant. The UV opacity is then calculated through:
\begin{equation}
 \qquad\qquad\qquad\qquad A_\mathrm{uv}=2.31(\beta-\beta_{0}),
\end{equation}
\\
\noindent where $\beta_{0}$ is the intrinsic UV spectral slope, fixed at $\beta_{0}=-2.1$ for $\beta>-1.4$ or $\beta_{0}=-2.35$ for lower measured UV slopes, as prescribed in \citet{Calzetti2000}. The mean $\beta$ values obtained for $z\sim4$, $z\sim5$ and $z\sim6$ are $-1.85\pm0.49$, $-1.98\pm0.57$ and $-2.19\pm0.66$, respectively. These $\beta$ values are consistent with the literature \citep[see, \textit{e.g.},][and references therein]{Dunlop2012}.\\

A problem with the Kennicutt method is that the Ly$\alpha$ line is very sensitive to both HI resonant scattering and dust absorption. Therefore the actual Ly$\alpha$ emission of the galaxy can be greatly affected. Indeed, the photon escape fraction is very uncertain \citep[see, \textit{e.g.},][]{Hayes2010}. On the other hand, the continuum is much less affected by absorption and scattering due to neutral gas so we assume that SFRs calculated using the rest-frame UV continuum luminosity at 1500 \AA{} are more reliable. In those cases where we have LBGs with emission line we can compare their SFRs calculated via one method and the other. In every case, the SFRs calculated using eq.~(\ref{eq:Madau}) are much larger, reaching three orders of magnitude difference for 11 candidates with bright $M_{1500}$ but a small Ly$\alpha$ line (2.7\% of the sample presenting both line emission and well measured UV continuum). This large difference can be due to: 1) the uncertainties in measuring the Ly$\alpha$ line extinction, and 2) the fact that the continuum measurement integrates not only the current starburst but also the previous history of star formation of that source. Moreover, for those LBGs with line emission, the ratio between the two SFRs seems to be stochastic, since, as we explain in detail in \citet{Arrabal2018letter}, the intensity of the line emission relative to the continuum heavily depends on the strength of the current starburst, which quickly decays within a few Myr. That said, the ratio between SFRs, measured with the Kennicutt and the Madau methods, will only depend on the current stage of the young starburst and the number of former recent starbursts undergone by a given source.\\

\begin{figure}
	\includegraphics[width=\columnwidth]{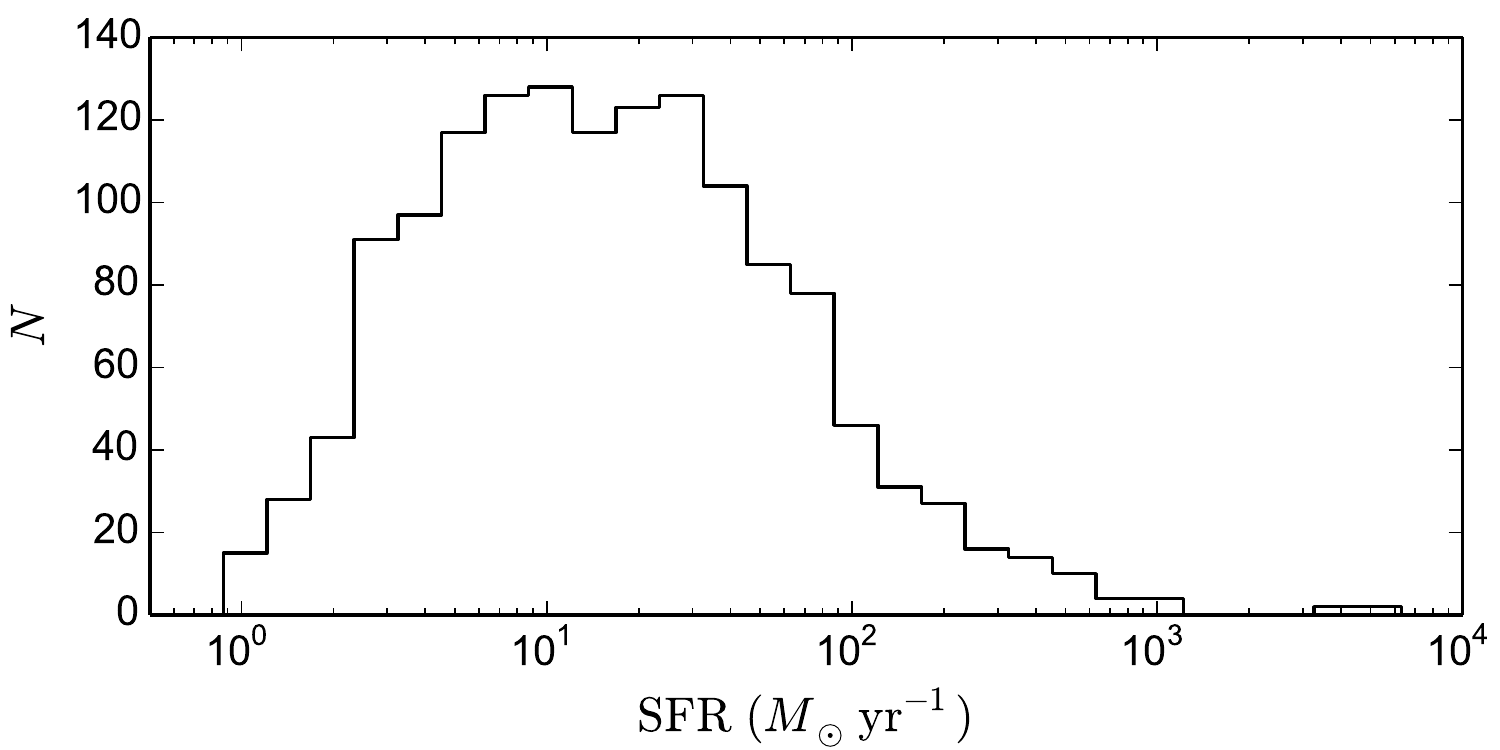}
    \caption{LBGs SFR histogram. Calculated through $L_{1500}$. The sample presents an average value of $\sim58\ M_{\odot}\ \mathrm{yr^{-1}}$, with seven objects above 1000 $M_{\odot}\ \mathrm{yr^{-1}}$.}
    \label{fig:SFR_distribution}
\end{figure}

\begin{figure}
	\includegraphics[width=\columnwidth]{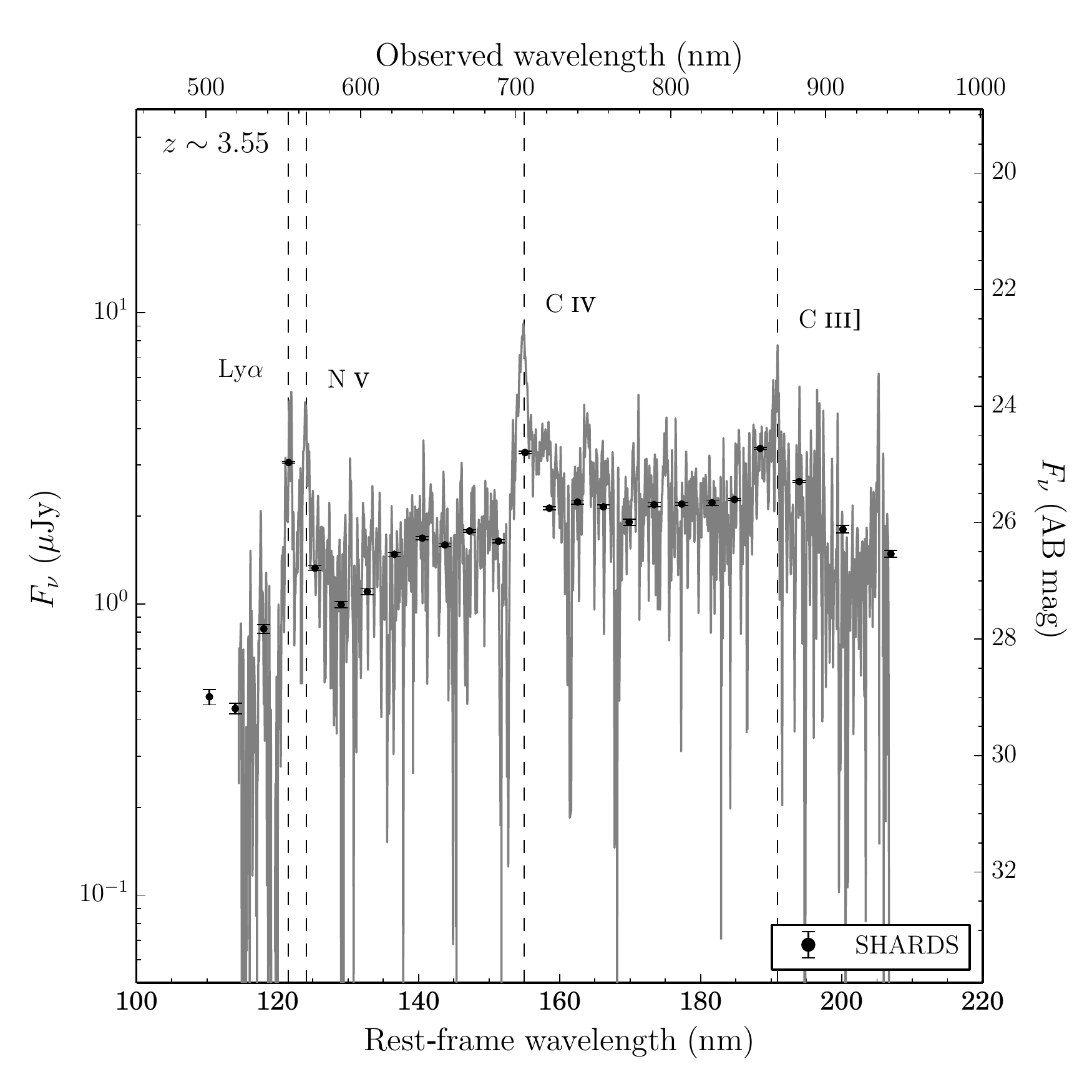}
    \caption{SED of the object SHARDS J123723.72+622113.0, a dusty galaxy with $\beta=-0.71\pm0.40$, resulting in $A_\mathrm{uv}=3.22\pm0.93$ mag and a dust corrected SFR of $1107\pm75$ $M_{\odot}\ \mathrm{yr^{-1}}$. Overlapped in grey, we show the spectrum available in the Team Keck Treasury Redshift Survey (TKRS) from \citet{Wirth2004}. Ly$\alpha$, \ion{N}{v}, \ion{C}{iv} and \ion{C}{iii}] lines are identified.}
    \label{fig:Dusty_Galaxy}
\end{figure}

The SFRs, corrected for extinction following the above prescription, are given in Table~\ref{tab:Results}, where we show the SFR based on $L(Ly\alpha)$ and/or $L_{1500}$ depending on whether the galaxy presents emission line or continuum emission. As expected for these early galaxies, the SFRs are fairly high \citep[\textit{e.g.,}][]{Bouwens2007, Bouwens2015, Mashian2016,Casey2017}, with those calculated with $L_{1500}$ being $\sim58\ M_{\odot}\ \mathrm{yr^{-1}}$ on average. The complete LBGs SFR distribution can be seen in Fig.~\ref{fig:SFR_distribution}, where there are 7 objects forming stars at a rate above 10$^{3}$ $M_{\odot}\ \mathrm{yr^{-1}}$. These galaxies are very dusty star-forming galaxies \citep[see][]{Casey2017} whose SEDs show a very red UV slope. An example is given in Fig.~\ref{fig:Dusty_Galaxy}.

\subsection{Equivalent Widths}
\label{sec:EW}
The EW of the Ly$\alpha$ line tells us the relative strength of the young star formation burst. It relates the line emission due to the current starburst to the continuum produced by the more evolved population of the galaxy. For every LBG with line emission we obtain the EW measuring the continuum either in the adjacent SHARDS filter or, if this filter has a large uncertainty, calculating a weighted mean of the fluxes in nearby filters. If it is not possible to get a continuum from the SHARDS filters, the continuum is obtained  from HST broad band data. To establish a minimum reliably detected line flux we impose that the filter measuring the line emission should be at least $1\sigma$ above the continuum flux level. Applying this, we reach a minimum measured rest-frame $\mathrm{EW}=5.14\pm4.70$ \AA{}.\\

\begin{figure}
	\includegraphics[width=\columnwidth]{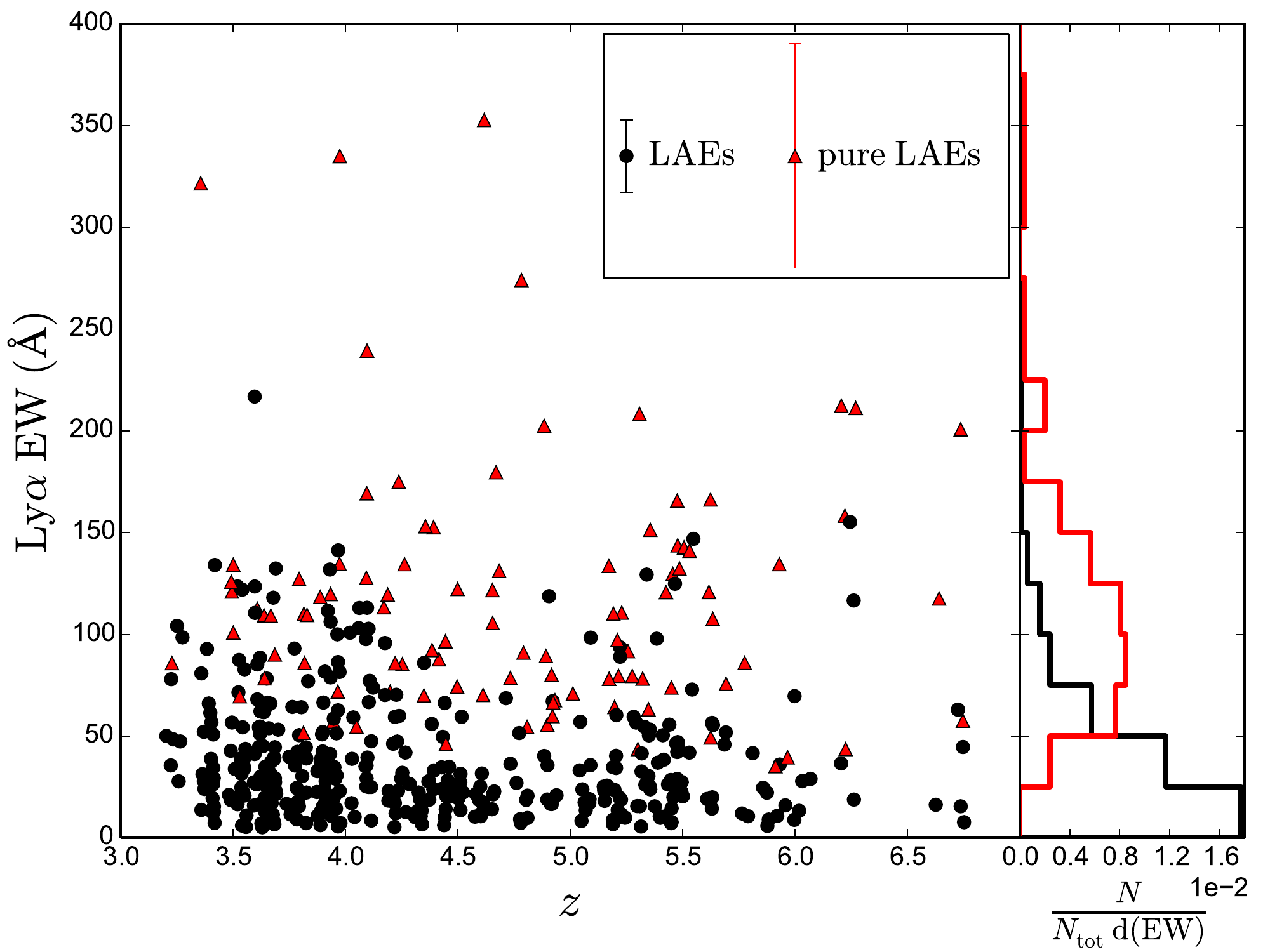}
    \caption{Rest-frame EW of the Ly$\alpha$ source candidates with emission line. Red triangles show the pure LAEs while the black circles are the LBGs with Ly$\alpha$ emission line. Error bars are omitted for convenience, but they are given in Table~\ref{tab:Results}. Median error bars are shown in the legend. The EW probability density distribution independent of $z$ is shown in the right panel. Pure LAEs show larger EWs than LBGs with emission line. Moreover, the number of sources increases at lower EWs for both populations.}
    \label{fig:EW}
\end{figure}

Measuring a continuum for pure LAEs is difficult as these objects are characterized by having very low or no detectable continuum emission. In some of the cases we can use broad band data but there are a few ones (28) where we cannot reliably measure their continua and therefore we cannot calculate the EW. As expected, pure LAEs have larger EWs than the rest of LAEs, although their uncertainties are also larger, since their continuum is much fainter or even absent. In our sample, the pure LAEs EWs are in the range 35-353 \AA{} with a median value of 109.13 \AA{}, while the LBGs/LAEs EWs are in the range 5.14-217 \AA{} with a median of 27.83 \AA{}. Nonetheless, the number of sources decreases as we move to higher EWs (see right panel in Fig.~\ref{fig:EW}). Notice that for the LBGs we plot only those 404 that are also LAEs, which means that there are 1030 additional LBGs with no emission line detected, \textit{i.e.}, with EW$<5.14$ \AA{}.\\

\begin{figure*}
	\includegraphics[width=\textwidth]{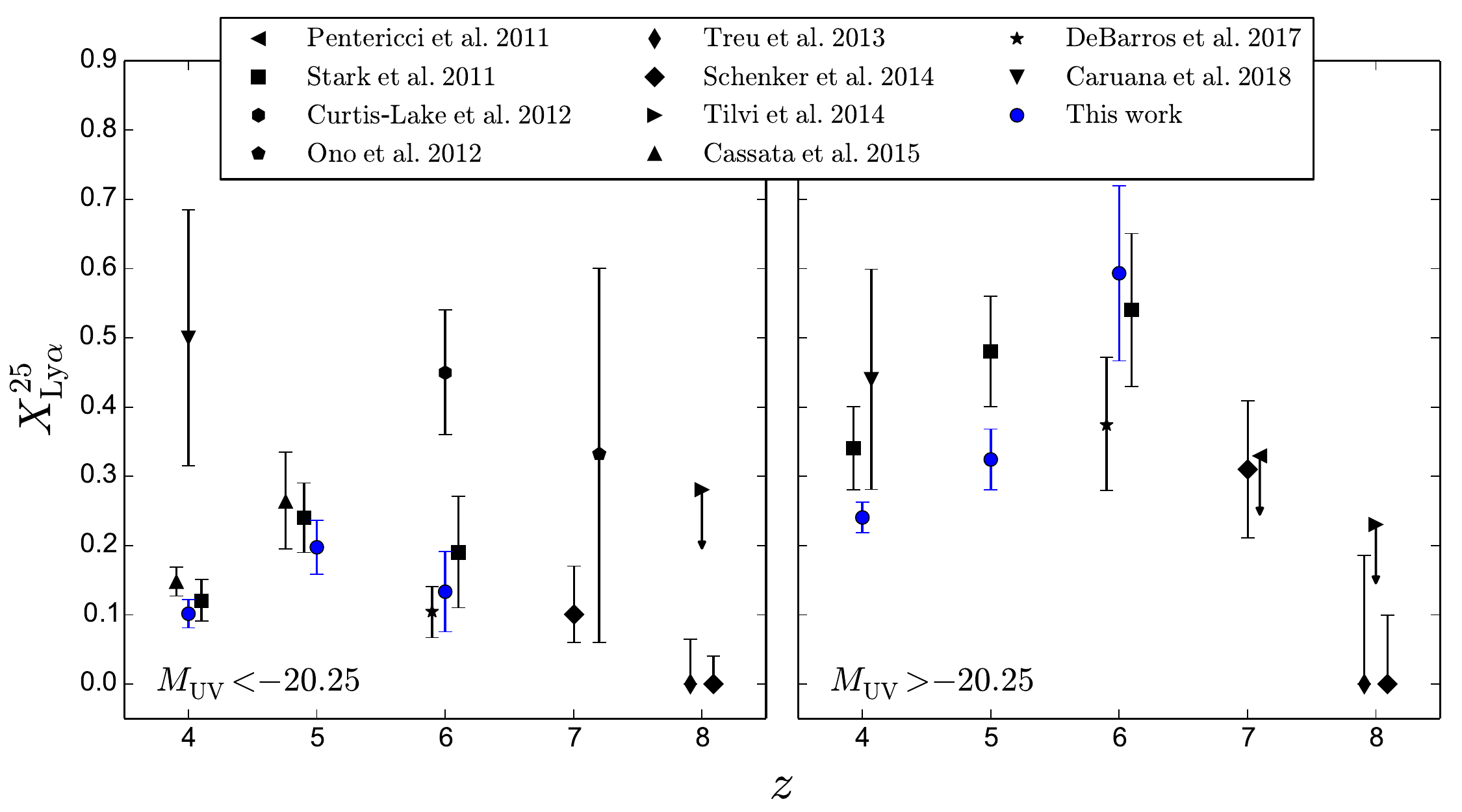}
    \caption{Fraction of objects with rest-frame Ly$\alpha$ EW$>25$ \AA{} versus  redshift. Left panel: galaxies brighter than $M_\mathrm{uv}=-20.25$. Right panel: galaxies fainter than $M_\mathrm{uv}=-20.25$. A slight offset in redshift is introduced to improve clarity. We show results from \citet{Pentericci2011}, \citet{Stark2011}, \citet{Curtis-Lake2012}, \citet{Ono2012}, \citet{Treu2013}, \citet{Schenker2014}, \citet{Tilvi2014}, \citet{Cassata2015}, \citet{De Barros2017} and \citet{Caruana2018}.}
    \label{fig:EW_25-Z}
\end{figure*}

In this work we have detected LAEs up to quite low EWs. However, in order to compare with previous studies, we have analysed the fraction of them with EW larger than a standard commonly used value of Ly$\alpha$ $\mathrm{EW}>25$ \AA{}. The comparison is done with various authors at different redshifts \citep{Pentericci2011,Stark2011,Curtis-Lake2012, Ono2012, Treu2013, Schenker2014, Tilvi2014, Cassata2015, De Barros2017, Caruana2018}. Results are shown in Fig.~\ref{fig:EW_25-Z}, where we divided the sample into two brightness groups. Our fraction of LAEs with $\mathrm{EW}>25$ \AA{} ($X_{\mathrm{Ly}\alpha}$) matches well those previously obtained, although we register a slightly smaller fraction of the faintest sources at $z\sim4$ and $z\sim5$. The decrease of $X_{\mathrm{Ly}\alpha}$ beyond $z\sim6$ found by other authors \citep[\textit{e.g.},][]{Caruana2014,Schenker2014} suggests an increase of the \ion{H}{i} abundance as we move to a non-fully reionized universe \citep[see, \textit{e.g.},][]{Fan2006}, which would strongly affect the scattering of the Ly$\alpha$ line.

\begin{table*}
	\centering
	\caption{Main relevant parameters of the sample: name of the object using the SHARDS identification, right ascension and declination, photo-redshift calculated, Ly$\alpha$ luminosity (in case the Ly$\alpha$ emission line is available), apparent magnitude derived at a rest-frame wavelength of 1500 \AA{} (not measurable in some pure LAEs), SFR derived from the Ly$\alpha$ emission, SFR derived from the rest-frame UV continuum at 1500 \AA{}, and rest-frame Ly$\alpha$ EW. The full version of this table is available in the on-line version.}
	\label{tab:Results}
	\begin{tabular}{ccccccccc}
		\hline
Object Name & R.A. & Dec. & $z$ & $L_{\mathrm{Ly}\alpha}$ & $m_{1500}$ & SFR$_{L_{\mathrm{Ly}\alpha}}$ & SFR$_{L_{1500}}$ & EW\\
            & (J2000) & (J2000) &  & ($10^{42}$ erg s$^{-1}$) & (AB mag) & ($M_{\odot}$ yr$^{-1}$) & ($M_{\odot}$ yr$^{-1}$) & (\AA{})\\
\hline  
SHARDS20010117 & 12:35:48.07 & 62:12:02.39 & 4.28$\pm$0.06 & - & 27.5$\pm$0.6 & - &  1.7$\pm$0.9 & -\\ 
SHARDS20007539 & 12:35:48.12 & 62:12:03.78 & 5.38$\pm$0.07 &  0.9$\pm$0.2 & 25.8$\pm$0.2 & 0.8$\pm$0.2 & 11.2$\pm$1.7 &  15$\pm$ 3\\ 
SHARDS20012481 & 12:35:50.89 & 62:11:58.49 & 5.69$\pm$0.06 & - & 26.2$\pm$0.1 & - & 10.7$\pm$1.3 & -\\ 
SHARDS20005927 & 12:35:51.53 & 62:12:16.49 & 3.22$\pm$0.07 & - & 26.4$\pm$0.2 & - & 27.4$\pm$4.9 & -\\ 
SHARDS20005405 & 12:35:51.63 & 62:12:12.66 & 4.03$\pm$0.07 & - & 25.6$\pm$0.2 & - & 36.8$\pm$7.9 & -\\ 
SHARDS20008074 & 12:35:52.15 & 62:11:20.83 & 5.53$\pm$0.07 &  0.5$\pm$0.2 & 26.2$\pm$0.1 & 0.4$\pm$0.2 &  8.0$\pm$1.0 &  42$\pm$31\\ 
SHARDS20008444 & 12:35:53.20 & 62:10:32.94 & 4.01$\pm$0.07 & - & 26.8$\pm$0.6 & - &  2.8$\pm$1.5 & -\\ 
SHARDS20010810 & 12:35:53.38 & 62:10:23.25 & 5.12$\pm$0.06 & - & 27.2$\pm$0.3 & - &  6.2$\pm$1.5 & -\\ 
SHARDS20005669 & 12:35:54.09 & 62:10:32.87 & 3.36$\pm$0.07 &  0.2$\pm$0.2 & 26.4$\pm$0.3 & 0.2$\pm$0.2 & 137.1$\pm$36.3 &  14$\pm$10\\ 
SHARDS20011405 & 12:35:54.26 & 62:10:18.83 & 5.37$\pm$0.07 & - & 25.9$\pm$0.2 & - &  9.9$\pm$2.0 & -\\ 
SHARDS20006420 & 12:35:54.43 & 62:10:33.80 & 3.88$\pm$0.06 & - & 26.0$\pm$0.3 & - & 37.5$\pm$11.3 & -\\ 
SHARDS20006258 & 12:35:54.54 & 62:12:14.59 & 3.48$\pm$0.06 &  0.3$\pm$0.2 & 25.7$\pm$0.1 & 0.3$\pm$0.2 & 12.9$\pm$1.8 &  21$\pm$12\\ 
SHARDS20013727 & 12:35:55.03 & 62:12:04.79 & 5.96$\pm$0.07 &  0.3$\pm$0.2 & 26.2$\pm$0.2 & 0.3$\pm$0.2 & 160.5$\pm$33.0 &  16$\pm$10\\ 
SHARDS20009009 & 12:35:55.19 & 62:11:25.40 & 3.89$\pm$0.06 & - & 26.7$\pm$0.3 & - & 16.7$\pm$5.1 & -\\ 
SHARDS20006827 & 12:35:55.65 & 62:10:19.00 & 4.28$\pm$0.06 & - & 26.2$\pm$0.2 & - &  5.5$\pm$1.2 & -\\ 
SHARDS20008662 & 12:35:55.79 & 62:10:28.94 & 4.17$\pm$0.07 & - & 26.9$\pm$0.6 & - &  2.9$\pm$1.6 & -\\ 
SHARDS20008870 & 12:35:55.83 & 62:12:32.05 & 4.19$\pm$0.07 & - & 27.1$\pm$0.5 & - &  4.0$\pm$2.0 & -\\ 
SHARDS20010887 & 12:35:56.17 & 62:11:45.41 & 5.14$\pm$0.06 & - & 26.2$\pm$0.3 & - & 76.2$\pm$23.6 & -\\ 
SHARDS20010975 & 12:35:56.63 & 62:11:43.25 & 5.40$\pm$0.07 &  0.2$\pm$0.2 & 26.8$\pm$0.5 & 0.2$\pm$0.2 & 10.7$\pm$5.4 &  11$\pm$ 9\\
...&...&...&...&...&...&...&...&...\\
		\hline
	\end{tabular}
\end{table*}

\subsection{Distances and grouping}
\label{sec:Grouping}
To find objects with close neighbours, we look for galaxies at the same redshift (within redshift errors) that are inside a 30 kpc radius circular region, assuming that objects farther away would not be part of the group, as considered, \textit{e.g.}, in \citet{Mundy2017}. The distance between objects in the group is calculated from their angular separation using the angular diameter distance at their redshift as it appears in \citet{Hogg1999}.\\

202 of the candidates are in 92 close groups of two or more galaxies, which represent $\sim13$\% of the total sample. In particular, we register, 79 pairs, 10 trios, 2 quartets and a sextet. The remaining sources are isolated. Previous works looking for pairs or very close groups at these redshifts find similar fractions \citep{Conselice2009, Mundy2017}.\\

\begin{figure}
	\includegraphics[width=\columnwidth]{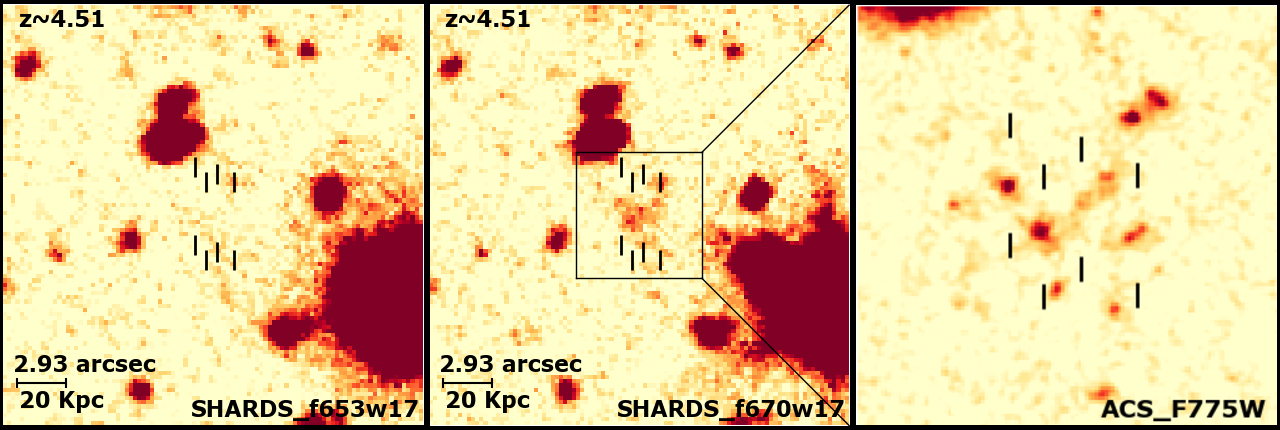}
    \caption{Images of the G6 quartet in two consecutive SHARDS filters spanning the Lyman break (f653w17 in the left panel and f670w17 in the middle one). North is up, East is left. Notice that the marked sources do not appear in the bluer filter. The right-hand panel shows a zoomed image of the objects in the F775W \textit{HST}/ACS image. A little plume can be appreciated in the westernmost source, which points to gravitational interaction between the objects of the group.}
    \label{fig:Cluster11}
\end{figure}

The distance between the objects in these groups vary from 59 kpc to 3.26 kpc. Three of the trios and one of the quartets are confined in a quite small area of around 20 kpc diameter. Many of the other trios are composed of a very close pair separated a larger distance from the third object of the group. Something similar happens with the sextet, which looks like four very close objects (one of which seems to be three different non-resolved sources) in a $\sim14$ kpc region, with the fifth and sixth ones somewhat more separated, showing a large and conspicuous tail between them. Table~\ref{tab:Distancias} lists the groups we found with their assigned names, indicating the distance between pairs or, in the case of groups with more members, the largest distance between the central object and the rest. The images of the quartet (G6) in the SHARDS bands sampling the Lyman break are shown in Fig.~\ref{fig:Cluster11}. We will discuss more about these close groups in Section~\ref{sec:Groups_and_Overdensities}.

\begin{table*}
	\centering
	\caption{Distances between the objects in groups and their redshifts. In case of more than two objects in a group, the distance shown is the largest between the central object and the rest. A full version of this table is available in the on-line version.}
	\label{tab:Distancias}
	\begin{tabular}{cccccc} 
		\hline
		    & Name           & R.A.(J2000) & Dec.(J2000)  & Distance (kpc) & Redshift\\
		\hline
		G1  & SHARDS10005314 & 12:37:24.06 & 62:18:33.45 &  6.57$\pm$0.09 & 3.42$\pm$0.07\\
		    & SHARDS10008105 & 12:37:24.01 & 62:18:32.67 &                & 3.42$\pm$0.07\\
		    & SHARDS10013381 & 12:37:24.10 & 62:18:34.25 &                & 3.42$\pm$0.07\\
		\hline
		G2  & SHARDS10004588 & 12:37:17.72 & 62:19:04.59 & 45.65$\pm$2.85 & 3.96$\pm$0.06\\
		    & SHARDS10007056 & 12:37:17.21 & 62:18:59.31 &                & 3.96$\pm$0.06\\
		\hline
		G3  & SHARDS20005310 & 12:36:28.03 & 62:09:01.82 &  5.86$\pm$0.08 & 3.92$\pm$0.06\\
		    & SHARDS20011181 & 12:36:28.14 & 62:09:01.52 &                & 3.92$\pm$0.06\\
		\hline
		G4  & SHARDS20005660 & 12:36:50.68 & 62:09:32.47 & 30.26$\pm$1.13 & 4.24$\pm$0.07\\
		    & SHARDS20009490 & 12:36:51.10 & 62:09:35.58 &                & 4.24$\pm$0.07\\
		\hline
		G5  & SHARDS10006493 & 12:37:44.84 & 62:18:17.25 & 20.56$\pm$0.73 & 5.09$\pm$0.06\\
		    & SHARDS10008700 & 12:37:44.98 & 62:18:20.28 &                & 5.09$\pm$0.06\\
		\hline
		G6  & SHARDS10009103 & 12:37:39.15 & 62:17:34.59 &  9.15$\pm$0.13 & 4.51$\pm$0.07\\
		    & SHARDS10007188 & 12:37:39.38 & 62:17:34.55 &                & 4.51$\pm$0.07\\
		    & SHARDS10007963 & 12:37:39.47 & 62:17:35.47 &                & 4.51$\pm$0.07\\
		    & SHARDS10014515 & 12:37:39.28 & 62:17:35.06 &                & 4.51$\pm$0.07\\
		\hline
		G7  & SHARDS20009174 & 12:37:01.39 & 62:09:09.95 & 50.23$\pm$4.24 & 4.35$\pm$0.06\\
		    & SHARDS20006656 & 12:37:01.04 & 62:09:16.78 &                & 4.35$\pm$0.06\\
		\hline
		G8  & SHARDS20006276 & 12:36:04.10 & 62:09:21.99 & 18.50$\pm$0.35 & 3.75$\pm$0.06\\
		    & SHARDS20005680 & 12:36:04.40 & 62:09:23.37 &                & 3.75$\pm$0.06\\
		\hline  
		...&...&...&...&...&...\\
		\hline  
	\end{tabular}
\end{table*}

\section{Discussion}
\label{sec:Discussion}
We have selected a high-$z$ star forming galaxy sample using the SHARDS survey as described in Section~\ref{sec:Database}. As shown before, our sample consists of 1434 LBGs and 124 pure LAEs, 202 of them forming pairs or close groups. Using the SHARDS data and all the extra information available in the Rainbow Database we have calculated 
redshifts, SFRs, comoving distances, Ly$\alpha$ EWs and $X_{\mathrm{Ly}\alpha}$ (Section~\ref{sec:Physical_parameters}). In this section we discuss the main results of our study.

\subsection{Comparison with previous studies}
\label{sec:comparison}
Keeping in mind our completeness magnitude, we compare our high-$z$ galaxy sample results with previous works. As reference, we took the big survey carried out by \citet{Bouwens2015}, hereinafter B15. In that work the LBGs search is made using HST broad band colour criteria. They find a much higher number of objects than we do. Indeed, in the GOODS-N field, B15 find 3917 LBGs, out of which 3455 fall within our redshift range according to their classification. Instead we only selected 1558 objects. To understand this discrepancy we analyse the B15 GOODS-N sample.\\

First of all we remove the GOODS-N B15 objects that fall out of our field of view. Their images survey 133.7 arcmin$^{2}$ in the GOODS-N field, while our effective search field is only $\sim128.4$ arcmin$^{2}$. Moreover, the centring and exact shape of the two fields differ slightly. We miss 405 galaxies (11.4\%) just because of this effect, leaving still 3050 objects to explain. However, only 2757 of them have photometric detection in at least one of the SHARDS filters. This means we are missing 293 sources, or 9.6\% of the B15 sample falling within our field, of which there are no detections in SHARDS.\\

\begin{figure}
	\includegraphics[width=\columnwidth]{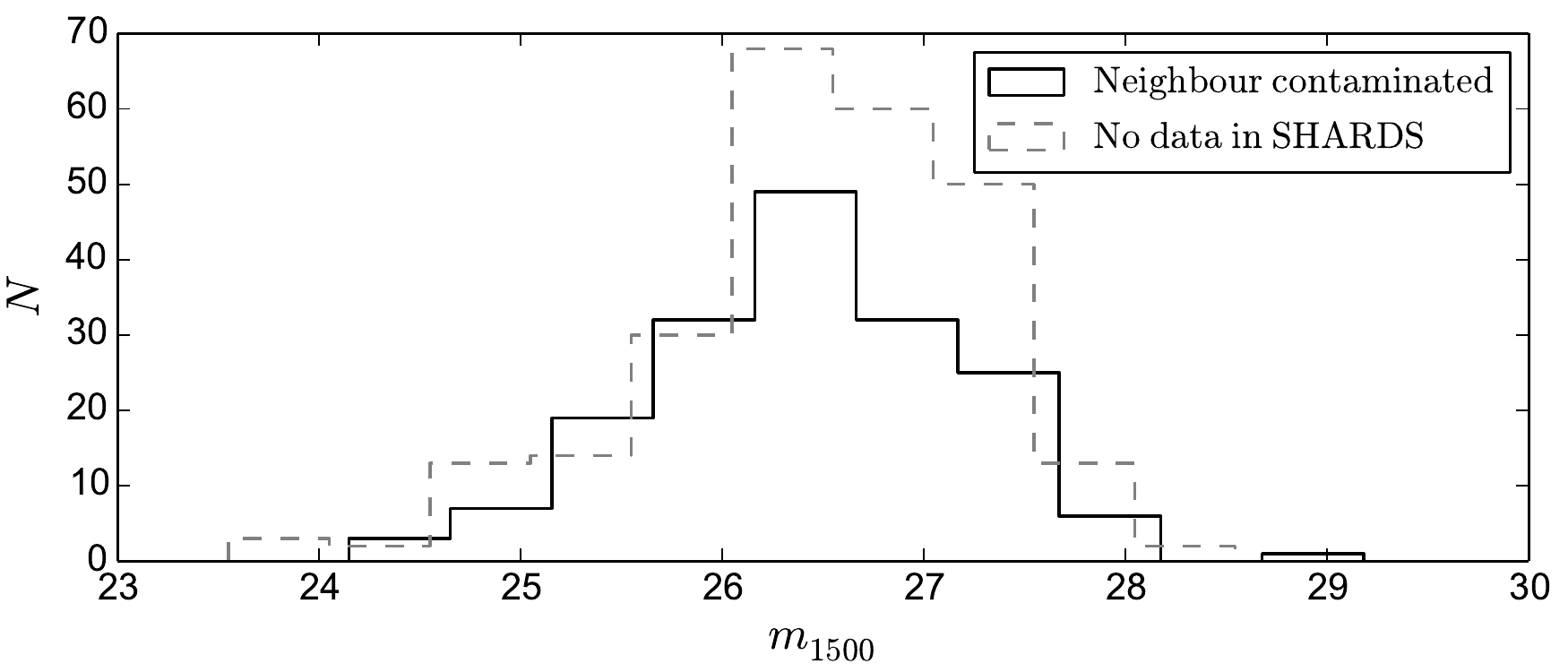}
    \caption{Rest-frame $m_{1500}$ distribution of the objects present in the B15 sample that are missed in this work because of non-detection or seeing issues. The $m_{1500}$ was measured using HST broad band photometry.}
    \label{fig:perdidos}
\end{figure}

Another effect to consider is the loss of sources due to neighbour contamination. This is so because of the higher spatial resolution of the HST in comparison to our ground-based GTC/OSIRIS resolution. To quantify this effect, we mask the stars and other big objects in the SHARDS images and analyse how many of the B15 candidates would be affected by them. For the smaller objects, we assume that any B15 galaxy with a brighter (or at most 0.5 mag fainter) object at a distance of $\theta=0.9$ arcsec or less (typical SHARDS seeing) is at least partially contaminated in our images, and therefore we could be missing it. 192 objects (7\%) are affected by this issue. The $m_{1500}$ distribution of the objects missed because of non-detection or neighbour contamination is shown in \ref{fig:perdidos}. At this point there are still 2565 galaxies of the B15 that should be detected.\\

We then limit the remaining B15 sample to those objects that should be in our selection. That is sources brighter than our completeness magnitude, within our field, not affected by proximity contamination and whose redshifts fall in the range we can measure with our 25 SHARDS filters. As we are going to exclude galaxies falling outside our completeness $m_{1500}$, we first calculate the wavelength corresponding to 1500 \AA{} rest-frame for each B15 object, which depends on $z$, to identify the HST filter that better samples that wavelength in each case. Then we look for every object in its corresponding filter in the \textit{HST}/ACS catalogue to check whether or not we should detect it with the SHARDS data. Up to $m_{1500}=25.87$ AB, there are 575 sources meeting all our limiting conditions, however, out of these, 115 (20.0\%) are not actually high-$z$ galaxies neither in our study nor according to the photometric redshifts from Ba18.\\

That said, we are conscious that the magnitude limit and spatial resolution in our search cause a loss of sources. Nevertheless, even considering these effects, we still find a non negligible fraction of objects in the B15 sample that meet our limiting magnitude condition but are not selected applying our criteria. These missing sources are visually revised in the SHARDS images to confirm they are bright enough and isolated enough not to be contaminated by any neighbour. We find that these sources do not look like LBGs when seen with our higher spectral resolution. Instead, these objects seem to be either lower $z$ Balmer break galaxies, cool stars or just red galaxies. Indeed, when these are seen through a few broad band filters only, they could be easily misclassified as LBGs, as they may present a reasonable flux drop from one broad filter to the next. We therefore conclude that, thanks to the much better wavelength discrimination of the SHARDS survey, we identify these objects as interlopers.\\

Note that we have not calculated photometric redshifts for objects not in our sample. However, as they have not been selected with our criteria, they are considered lower $z$ sources. In order to study the redshift distribution of these not selected sources, we make use of the redshifts calculated in Ba18. In addition, we repeat the analysis previously described in this section using different values of magnitude limit and flux contamination distance (seeing). The number of B15 galaxies that meet our limiting conditions but whose SHARDS photometric $z$ is much lower than that given in B15 is shown in Table~\ref{tab:Interlopers_Bouwens}, for different values of $m_{1500}$. In this Table, we use both 0.9 arcsec, corresponding to the typical SHARDS seeing, and 1.2 arcsec, a bit above that seeing, as the minimum distance between objects at which they can be considered isolated. There is little change in the interloper fraction for limiting magnitudes up to $\sim26.13$ AB, mostly constant at $\sim20$\%. Beyond that, this fraction slightly increases (up to 22.3\% at $\sim27.5$ AB), which makes sense since fainter objects are more difficult to characterize and identify as either good galaxy candidates or interlopers.\\

\begin{table}
	\centering
	\caption{Number of broad band selected objects from the B15 sample detected in SHARDS with different brightness and isolation conditions and the number of them that are considered lower $z$ interlopers according to the SHARDS photometric fits. The redshift distribution of these objects can be seen in Fig.~\ref{fig:hist_interlopers}.}
	\label{tab:Interlopers_Bouwens}
    \begin{threeparttable}
	\begin{tabular}{ccccc} 
		\hline
		$m_{1500}$ & $\theta$\tnote{1}   & N in B15 & N in B15      & Interlopers\\
		limit      & (arcsec) & (total)\tnote{2}  & (interlopers)\tnote{2} & fraction (\%)\\
		\hline
		25.50     & 0.9      & 329      &   68         & 20.67\\
		25.50     & 1.2      & 311      &   55         & 17.68\\
		25.87     & 0.9      & 575      &   115         & 20.00\\
		25.87     & 1.2      & 539      &   92         & 17.07\\
        26.00     & 0.9      & 666      &   133         & 19.97\\
		26.00     & 1.2      & 628      &   108         & 17.20\\
		26.13     & 0.9      & 762      &   151         & 19.82\\
		26.13     & 1.2      & 721      &   126         & 17.48\\
		26.50     & 0.9      & 1133      &  234          & 20.65\\
		26.50     & 1.2      & 1074      &  201          & 18.72\\
		27.00     & 0.9      & 1674      &  359          & 21.45\\
		27.00     & 1.2      & 1582      &  316          & 19.97\\
		27.50     & 0.9      & 2077      &  463          & 22.29\\
        27.50     & 1.2      & 1959      &  416          & 21.24\\
		\hline

	\end{tabular}
    \begin{tablenotes}
     \item[1] Isolation distance around a source to consider it free from contamination by neighbours in the SHARDS images.
     \item[2] Only objects with assigned redshift in B15 within our range of study.
    \end{tablenotes}
    \end{threeparttable}
\end{table}

\begin{figure}
	\includegraphics[width=\columnwidth]{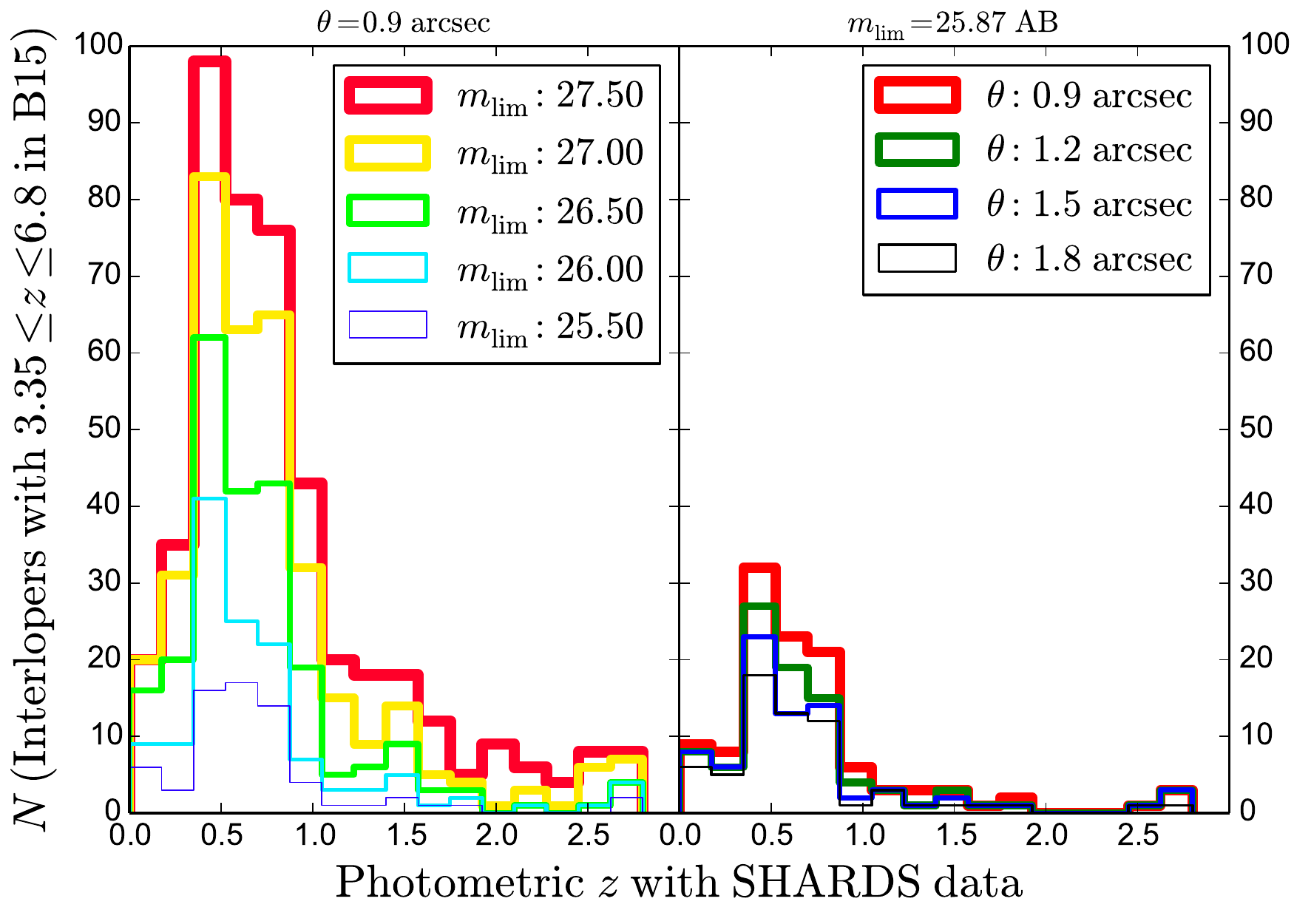}
    \caption{Redshift distribution of the objects in the B15 sample that appear as interlopers according to both our selection criteria and the photometric fits with SHARDS data from Ba18. On the left, different colours correspond to different $m_{1500}$ limits with a fixed isolating distance (0.9 arcsec). On the right, different isolating distances with a fixed $m_{1500}$ limit of 25.87 AB mag, corresponding to our 90\% completeness. The shape of the distribution is practically the same in all cases. The peak around $z\sim0.5$ suggests a wrong identification of the Balmer break as the Lyman break.}
    \label{fig:hist_interlopers}
\end{figure}

\begin{figure}
	\includegraphics[width=\columnwidth]{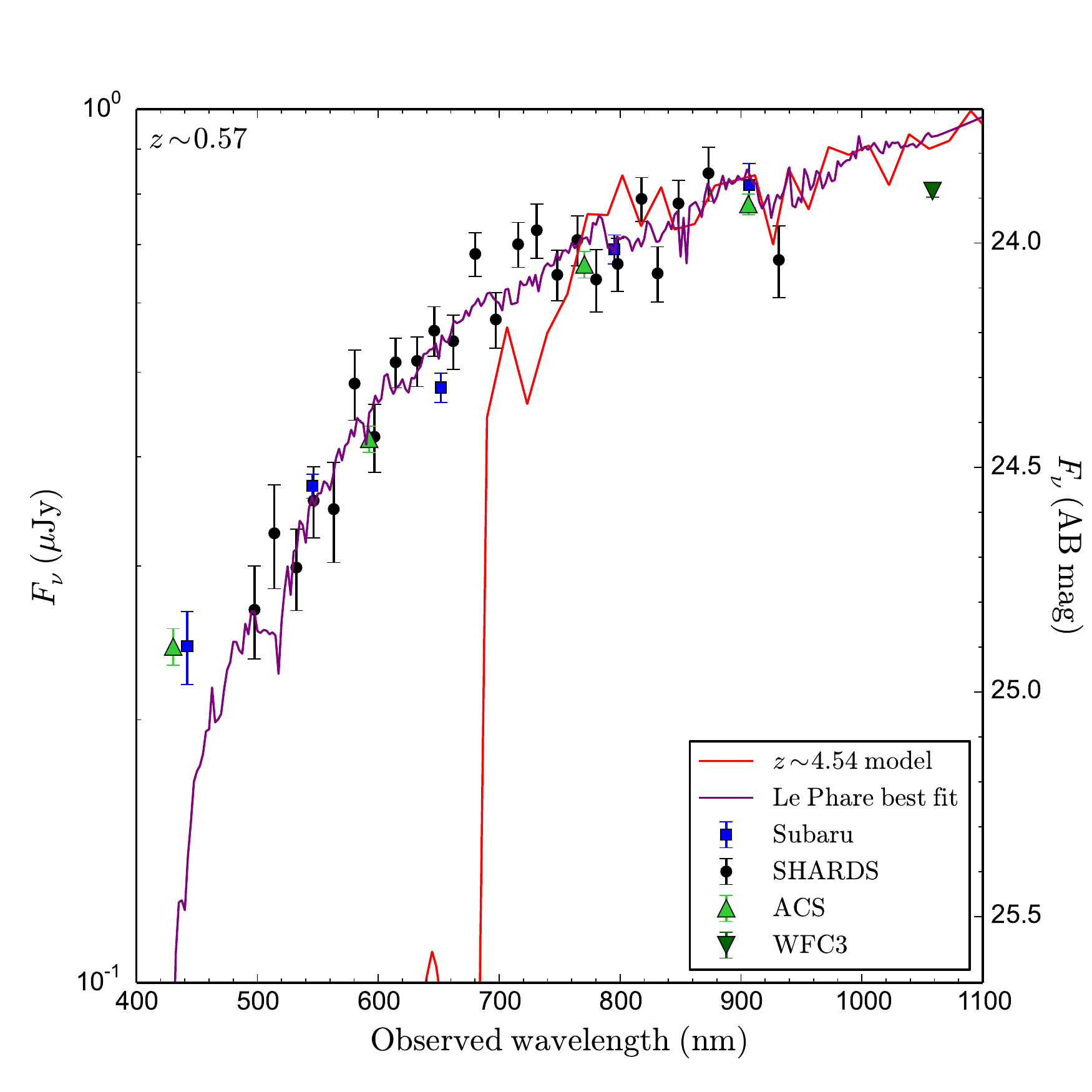}
    \caption{SED of the object SHARDS J123659.36+621518.7, wrongly selected as a $z\sim4.54$ LBG using broad band criteria. The information as given by the green triangles (ACS and WFC3 data) can lead to an incorrect Lyman break detection. A \textit{Le Phare} fit with the complete SHARDS data (purple line) shows that this is a $z\sim0.57$ object. The red line shows where the break would appear if this were a $z\sim4.54$ galaxy.}
    \label{fig:mal_clasificado}
\end{figure}

Therefore we conclude that $\sim20$\% of the B15 sample, up to $m_{1500}\sim25.87$ AB, is actually misclassified as high-$z$ LBGs \citep[quite above the $\sim2$-6\% found by][]{Vulcani2017}. It is interesting to see the redshift distribution of these misidentified objects. Figure~\ref{fig:hist_interlopers} is a histogram representing them as a function of redshift for different $m_\mathrm{lim}$ and isolation distances. We find a clear concentration of sources at very precise redshifts. Moreover, the shape of the interlopers redshift distribution is practically the same independently of the depth and seeing considered. This implies that these interlopers are not due to seeing or brightness issues, but to the spectral resolution quality. In fact, the distribution peaks around $z\sim0.5$, and a big fraction ($\sim$70\%) of the misclassified sources are distributed between $z=0.3$ and $z=0.9$, corresponding to a range between $z\sim3$ and $\sim5$ if the Balmer break is mistakenly taken as the Lyman break. We would actually expect a similar fraction of interlopers in other broad band studies with similar colour selection criteria. Fig.~\ref{fig:mal_clasificado} shows an example of one of these misidentified objects selected in B15.\\

\begin{figure}
	\includegraphics[width=\columnwidth]{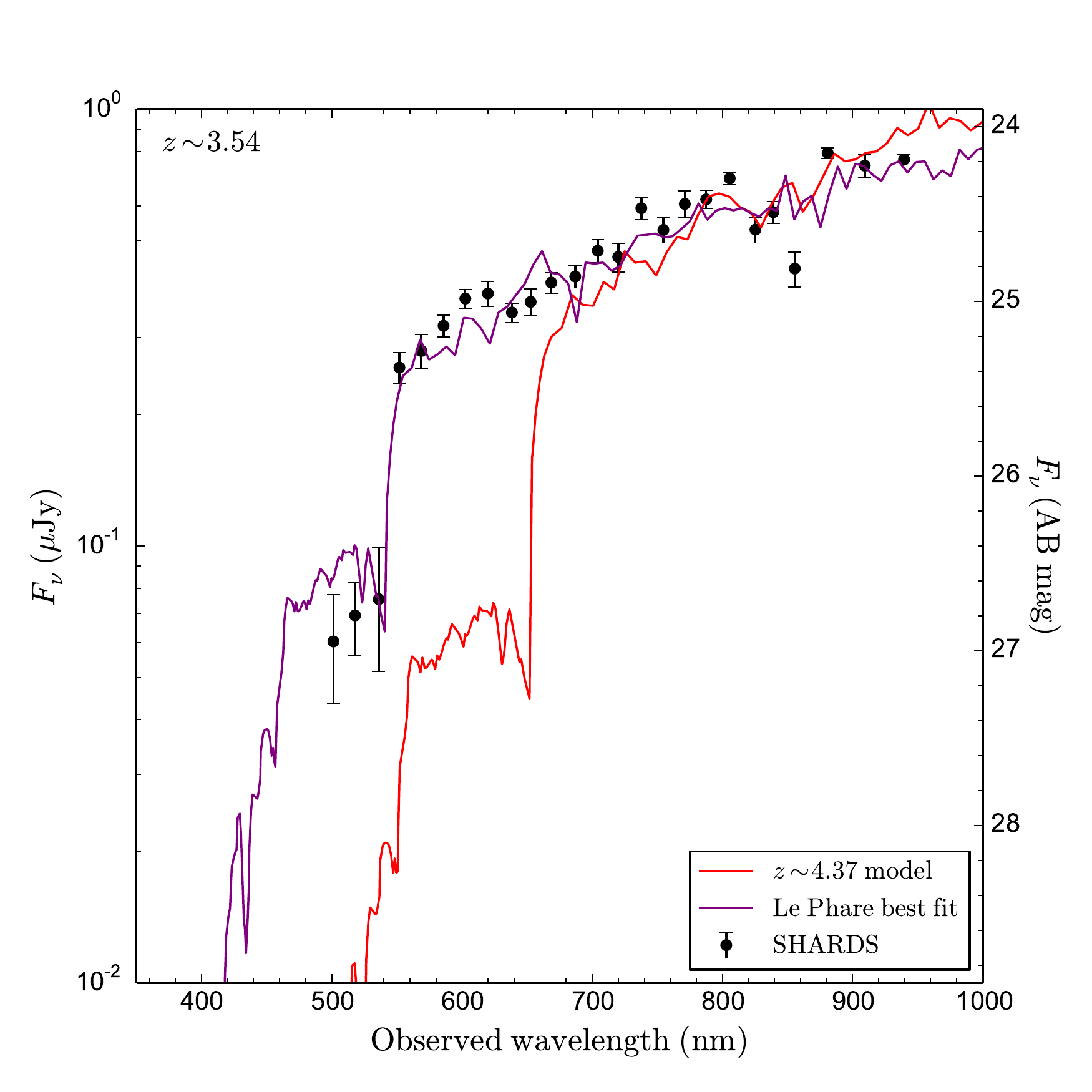}
    \caption{SED of one of our LBGs at $z\sim3.54$, SHARDS J123643.53+621121.4, wrongly selected as an object at $z=4.37$ using broad band criteria. At that redshift we should see the Lyman Break around 640 nm, which is clearly not the case when observed with the higher SHARDS resolution. Our $z$ matches fairly well the \textit{Le Phare} fit (purple line). The red line shows where the Lyman Break would be if it were at $z=4.37$.}
    \label{fig:mal_z}
\end{figure}

Finally, for some galaxies in the B15 sample that also meet our selection criteria, the redshift assigned in B15 does not match ours within a $\Delta z=0.3$ error (approximately twice our highest photo-$z$ error). An example of this is shown in Fig.~\ref{fig:mal_z}. To better characterize this discrepancy, we study the subsample of objects with photo-$z$ calculated both in B15, Ba18 and this work, consisting on 933 galaxies. The B15 redshift of 138 of these sources ($\sim15$\%) are found to differ from both Ba18 and our calculated redshifts. We consider our redshift determination much more accurate since we have a much better spectral resolution in our data, and our redshifts are in an excellent agreement with spectroscopic values as shown in Fig.~\ref{fig:Zspec_Zphot}. Indeed, the photo-$z$ obtained with broad band filters show a considerable uncertainty.\\

\begin{figure}
	\includegraphics[width=\columnwidth]{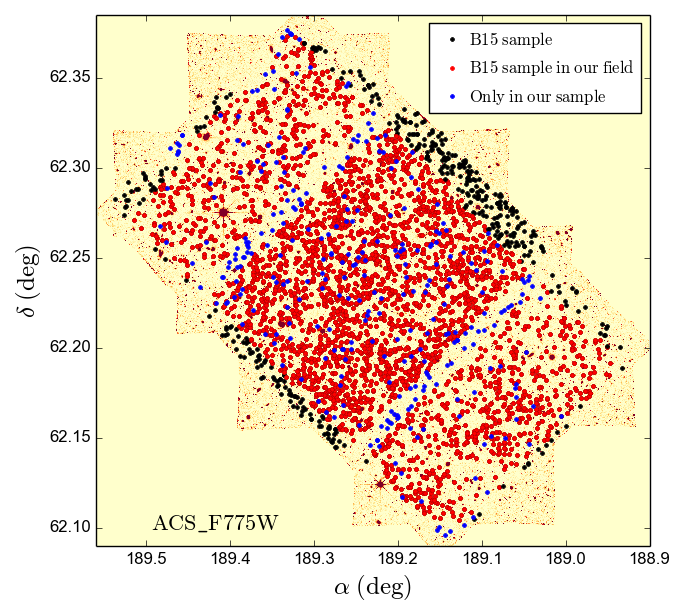}
    \caption{Spatial differences between the GOODS-N field sampled in B15 and the one sampled with SHARDS. The black dots are included in the B15 sample, but not in ours. The red dots are those within our SHARDS effective field. The blue dots are objects selected in our sample but not in B15. Small gaps of the B15 GOODS-N field contain 121 of them, but there are still 191 more within their effective field.}
    \label{fig:diferencias_campos}
\end{figure}

\begin{figure}
	\includegraphics[width=\columnwidth]{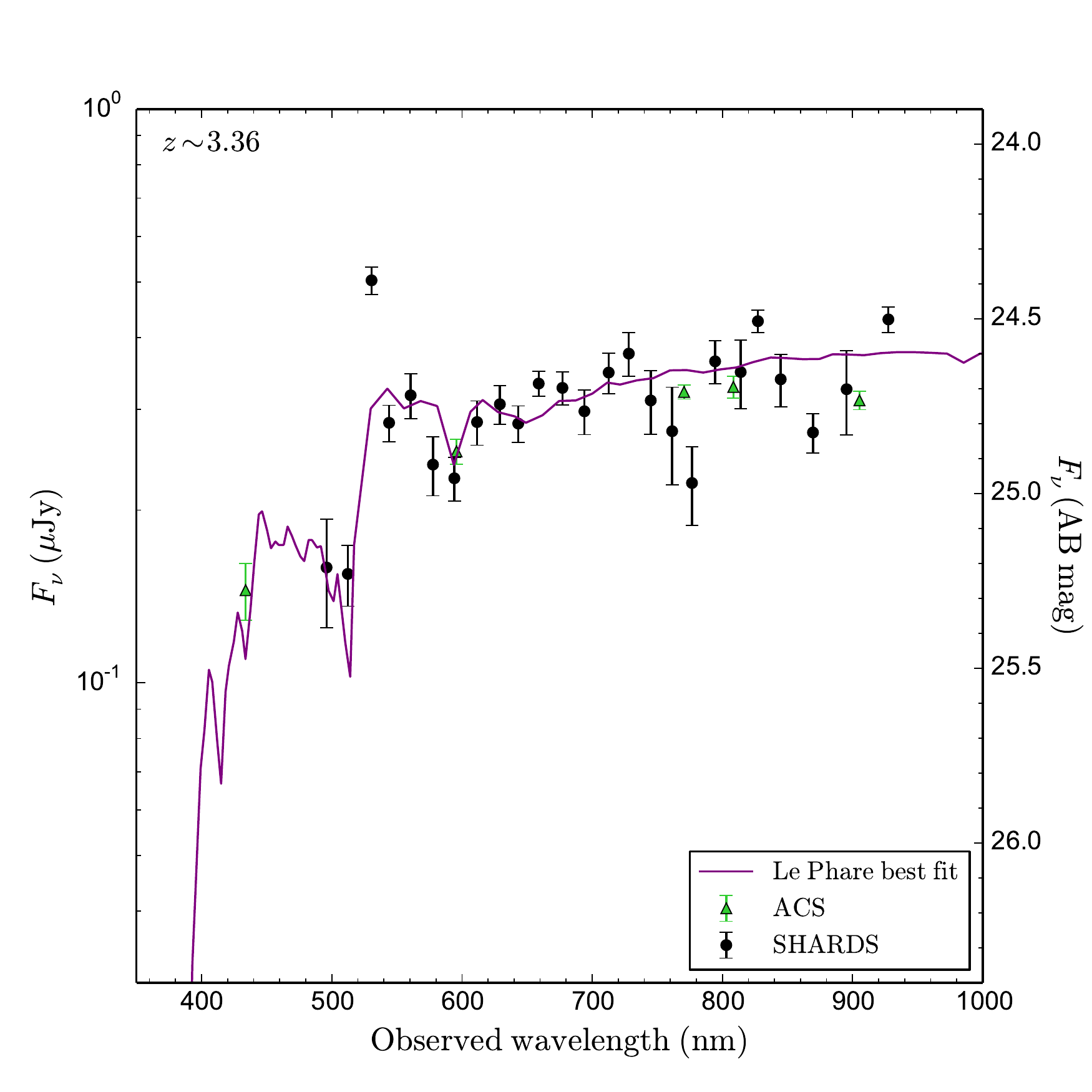}
    \caption{SED of the object SHARDS J123646.68+621517.1, selected with our criteria but not with the broad band selection criteria used in B15. The best \textit{Le Phare} fit is shown in purple.}
    \label{fig:solo_nuestro}
\end{figure}

Interestingly, there are other 312 objects in our sample that are not selected by B15. Of them, 121 are placed in two little gaps of the B15 GOODS-N field, as shown in Fig.~\ref{fig:diferencias_campos}. Of the remaining 191, 17 have spectroscopic $z$ available in Rainbow and/or NED that matches the photo-$z$ calculated in this work. A closer look to their SEDs does not give us any conclusive hint about why they are not selected using broad band filters. An example of these objects is shown in Fig.~\ref{fig:solo_nuestro}. Some of these galaxies are pure LAEs with strong line emission and a very faint continuum that could have been missed with broad band filters. This explanation is however valid for just 16 objects out of the 191. We do not know why the rest are not selected with the broad band criteria.\\

Summarizing, from the 3455 objects detected by B15 in GOODS-N within our $z$ range, after considering limiting brightness difference, small field of view differences and possible neighbouring light contamination, we would expect to detect 575 of them up to our 90\% completeness magnitude. Of them, 115 are not selected and are classified as low $z$ objects, implying that $\sim20$\% of the B15 broad band selected sample up to $m_{1500}=25.87$ AB mag seem to be interlopers. On the other hand, we have further found 191 galaxies that we consider good candidates and were not selected in B15.

\subsection{Luminosity Functions}
\label{sec:LF}
\begin{figure}
	\includegraphics[width=\columnwidth]{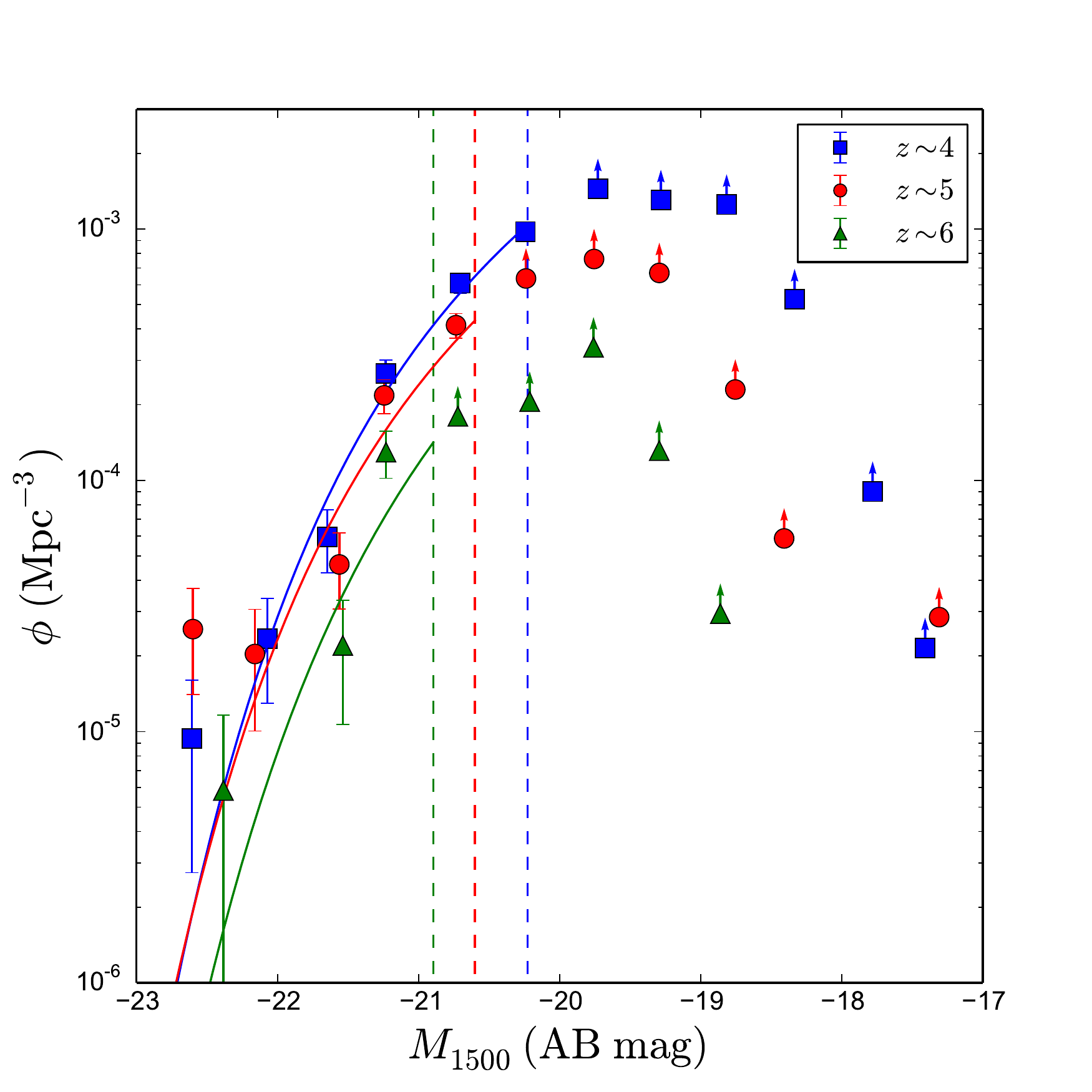}
    \caption{LFs of our LBG sample divided in three $z$ ranges: $z\sim4$, $z\sim5$ and $z\sim6$. The solid lines are the best-fit obtained for the exponential term of a Schechter function for each $z$, while the dashed vertical lines show 90\% completeness absolute magnitude for each redshift.}
    \label{fig:LF_Breaks}
\end{figure}

We proceed now to build the LBGs LFs, which are shown in Fig.~\ref{fig:LF_Breaks}. Our 90\% completeness absolute magnitude limit for each redshift is marked with a dashed vertical line. A correction for those objects that we miss, either because of non-detection in the SHARDS images or because of neighbouring light contamination of their SEDs, is considered for each $z$ range, based on their $m_{1500}$ distribution (see Fig.~\ref{fig:perdidos}). The magnitude values are measured using HST photometry. To correct our data in the faint region of the LF, we use the $V/V_{\mathrm{max}}$ correction to adequately consider the effective volume sampled by each detected source. We divide our sample in three main redshift bins to facilitate comparing with the literature, namely $z\sim4$ ($3.5\leq z<4.5$), $z\sim5$ ($4.5\leq z<5.5$) and $z\sim6$ ($5.5\leq z<6.5$). A Schechter function is used to fit the data:

\begin{equation}
 \phi(M)=\phi^{*}\frac{\mathrm{ln}(10)}{2.5}\times \mathrm{exp}[-10^{0.4(M^{*}-M)}]\times10^{0.4(M^{*}-M)(\alpha+1)}.
\end{equation}
\\
Since we are missing a fair number of the faintest objects, we cannot get information about the power-law part of the Schechter function. Therefore, the $\alpha$ parameter cannot be directly calculated, so we fix its value to a value consistent with the literature. In particular, we set it at each redshift to the mean value obtained from the following authors: \citet{Bouwens2007}, \citet{vdB2010} and B15 at $z\sim4$, \citet{Bouwens2007}, \citet{Iwata2007}, \citet{McLure2009}, \citet{vdB2010} and B15 at $z\sim5$ and \citet{Bouwens2007}, \citet{McLure2009}, \citet{Bouwens2012}, \citet{Bowler2015} and B15 at $z\sim6$. It is important to note that the $\alpha$ values derived from the literature may be overestimated according to the large interloper fraction we find in typical surveys made through broad band observations. The bright region can be fitted with the exponential term of the Schechter function to estimate $\phi^{*}$ and $M^{*}$. The results obtained are listed in Table~\ref{tab:Schechter_params}. Furthermore, we notice that the brightest points of our LFs tend to be slightly over their fitted Schechter function. This is in agreement with recent big field studies that propose a power-law behaviour for the bright region of the LFs, instead of the classical Schechter function \citep{Sobral2017,Sobral2017b}. Nonetheless, our field is not large enough to claim conclusive results in this regard.\\

A cosmic variance calculator, developed by \citet{Trenti2008} using halo models, is also considered to evaluate the effect of cosmic variance and Poisson counts errors, as done \textit{e.g.}, in B15, \citet{Conselice2016} and \citet{Vulcani2017}. We thus calculate that the number counts at $z\sim4$, $z\sim5$ and $z\sim6$ has a fractional error due to cosmic variance of 16\%, 20\% and 28\%, respectively. These uncertainties are relatively large in our study as we are limited to a relatively small field.\\

\begin{figure}
	\includegraphics[width=\columnwidth]{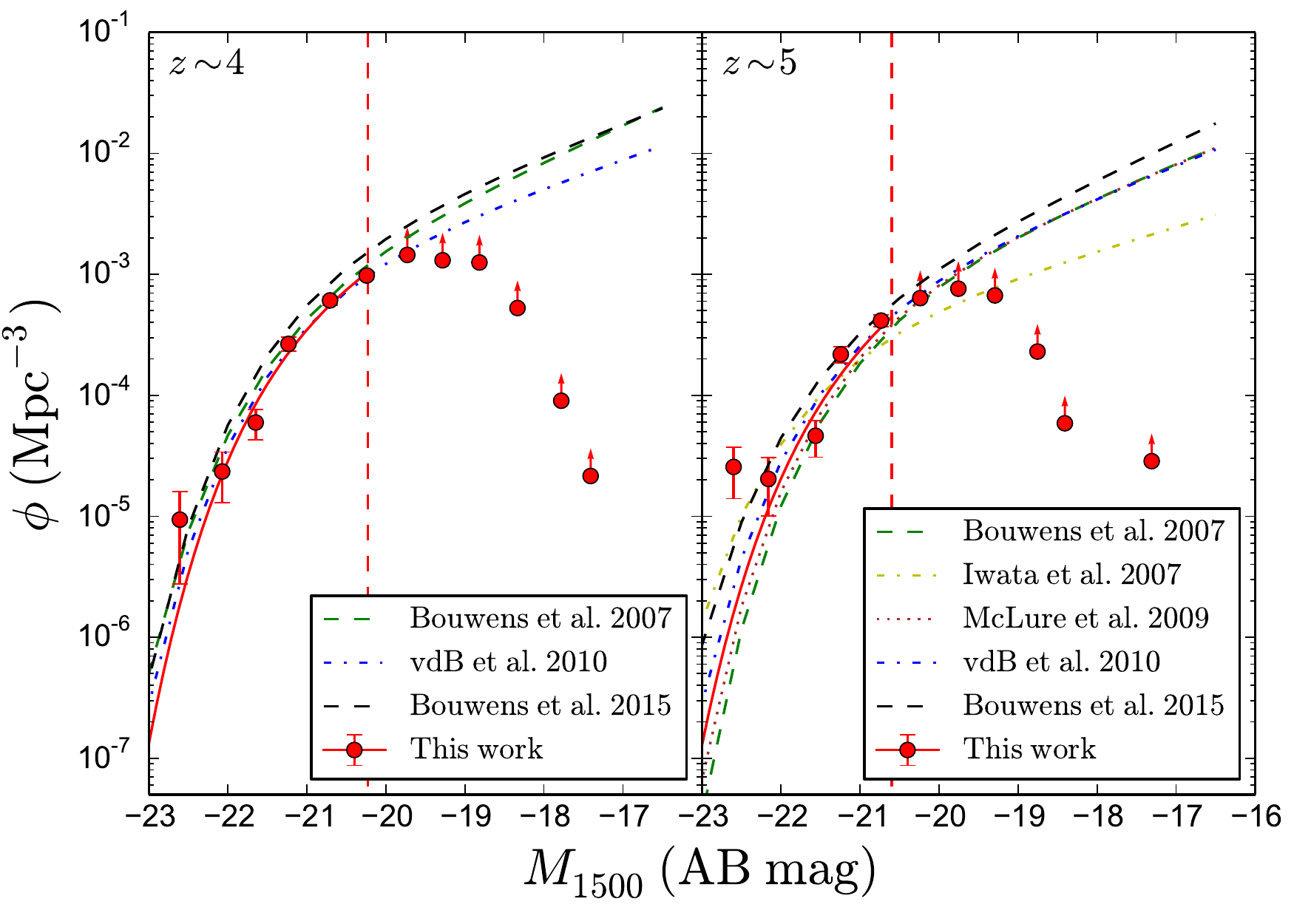}
    \caption{LFs of the most populated redshift ranges of our study ($z\sim4$ and $z\sim5$) overlaid with LFs from previous works. The red dashed vertical line indicates the 90\% completeness magnitude. The number of galaxies detected in this work is much lower than the numbers detected in previous ones with HST (see discussion in Section~\ref{sec:comparison}). 
    Our LFs fit well previous studies at the brightest region. However, they grow slower as we move to fainter magnitudes. 
    Our survey lacks the capability of sampling the faintest region of the LF but we expect it to be below previous values, due to interlopers issues in broad band studies.}
    \label{fig:LF_prev}
\end{figure}

At $z\sim6$ we have large uncertainties, due to the scarce statistics, cosmic variance effects and poissonian errors. For the other two redshift ranges, the results are comparable to previous studies (see Fig.~\ref{fig:LF_prev}). Notice that both at $z=4$ and 5, our LF estimations fit the bright part fairly well, decaying quickly as we move to fainter magnitudes.\\

\begin{table}
	\centering
	\caption{Schechter parameters for our LFs at $z\sim4$, $z\sim5$ and $z\sim6$.}
	\label{tab:Schechter_params}
    \begin{threeparttable}
	\begin{tabular}{cccc} 
		\hline
		$z$ & $M_{1500}^{*}$ & $\phi^{*}$ $(10^{-3}\ \mathrm{Mpc^{-3}})$ & $\alpha\tnote{1}$\\
		\hline
		4 & -20.75$\pm$0.13 & 1.55$\pm$0.31 & -1.64\\
		5 & -20.85$\pm$0.26 & 0.90$\pm$0.34 & -1.64\\
		6 & -20.74$\pm$0.58 & 0.54$\pm$0.61 & -1.79\\
		\hline  
	\end{tabular}
    \begin{tablenotes}
     \item[1] Fixed to the mean value given in the literature referred in the text.
    \end{tablenotes}
    \end{threeparttable}
\end{table}

We do match previous studies in the bright region. However, we find a discrepancy in the intermediate magnitudes region. We are indeed missing candidates because of seeing contamination in this region, but this effect is already corrected, though not enough to explain the difference we find, since the seeing contamination only affects a $\sim7$\% of the cases, as described in Section~\ref{sec:comparison}. On the other hand, the reliability of our candidates is very high, suggesting that the reason for the discrepancy is not only due to a lack of faint objects in our sample, but to an overestimation in the detection of LBGs when the selection is made with broad band filters. As we have seen before, the number of interlopers in broad band surveys is large and increases with magnitude. This is a possible reason for the increasing deviation of our LFs as we move to fainter magnitudes. We cannot estimate the precise effect that the interloper fraction produces at magnitudes beyond our completeness limit, but we expect it to increase, implying an uncertain calculation of the Schechter $\alpha$ parameter in surveys done with broad band photometry. We would like to stress the importance of having good spectral resolution for building a high-$z$ galaxy sample free from interlopers. 
As an estimation, we would expect the ``real'' LF to be somewhere between ours and that given in previous HST works, to account for the seeing and limiting magnitude difference. If this study is correct, previous broad band LFs should be downscaled by a factor $\sim20$\%, which could be even more important at fainter magnitudes. This would presumably lead to a decrease in the faint end slope of the LFs.

\subsection{Pair, Groups and Overdensities}
\label{sec:Groups_and_Overdensities}
As commented in Section~\ref{sec:Grouping}, 92 groups are found within our sample. Almost half of them (48) are pairs separated by longer distances than the rest, though they could still be interacting sources according to the typical distances given in \citet{Duc2014}. In the other 44 groups, the galaxies are separated by distances of around 30 kpc (3.90-4.72 arcsec, depending on redshift) or less. For these very close sources it is difficult to distinguish whether they are independent galaxies or star-forming knots in a single clumpy galaxy \citep{Elmegreen2013}. Fitting their SEDs with simple models results in a mass of the order of 10$^{8}$ $M_{\odot}$, in which case they should be high-$z$ galaxies rather than clumps. The typical size of these objects is $\sim0.4$ arcsec, which is consistent with typical sizes of high-$z$ galaxies as found by \citep{Bouwens2004, Oesch2010, Ono2013, Shibuya2015, Holwerda2015, Liu2017}. Moreover, luminous clumps in clumpy galaxies are usually surrounded by a common isophote, embedding the entire galaxy, below the bright level of the clumps but above the background sky \citep[see][]{Hinojosa2016}. This effect is not seen in our objects. Because of these reasons we claim that these objects should actually be individual galaxies in compact groups.\\

In particular, the group G5 is an already spectroscopically confirmed pair of galaxies at $z\sim5.1$ in which the northern source presents a nebulosity around it that could be either a tidal tail caused by the interaction with the other galaxy  or the remnant of a recent merger with a third object \citep{J&C2014}. We find similar tail-like morphologies in most of our groups (\textit{e.g.,} the westernmost object of G6 shown in Fig.~\ref{fig:Cluster11}). These structures support the hypothesis that our groups are gravitationally bounded systems. In addition to the tidal tails we also notice some other little spots near the trios and one more near G6 in the \textit{HST}/ACS images that could be additional members of their respective groups. However, these extra objects do not appear either in the SHARDS or the ACS catalogues nor in the Rainbow database, and they cannot be resolved in the GTC images, so we do not have conclusive photometric information for them. In fact, some of the objects forming these groups seem to be more than one unresolved source.\\

Note that the number of groups found, though significant, is not large enough to warrant studies of their dependence on redshift. They seem to follow a similar $z$ distribution as the whole sample. Concerning their SFR, we check whether they present particularly high ones. We find that they have similar SFR distribution as the rest of the sources in the sample.\\

In addition to these very close groups, we have also detected 87 and 55 objects possibly belonging to two already known overdensities at $z=4.05$ \citep{Daddi2009} and $z=5.198$ \citep{Walter2012}, respectively, in the GOODS-N field. Given our small typical redshift errors ($\Delta z\sim0.07$) we do suggest they belong to those overdensities. This increases by a factor of $\sim4$ the number of sources previously found for the $z=5.198$ overdensity. If confirmed, this proto-cluster at $z\sim5.2$ will be one of the richest proto-clusters beyond $z=5$.

\section{Conclusions}
\label{sec:Conclusions}
We have used the SHARDS survey to carry out a search of high-$z$ galaxies. The narrow/medium width of the 25 SHARDS filters and their completeness in the 500-941 nm wavelength range have allowed us to develop very precise SEDs in the optical/NIR. We have therefore sampled the high-$z$ galaxy population from $z\sim3.35$ to $z\sim6.8$ in a uniform way. 
The special characteristics of the filters have allowed us to simultaneously detect both LAEs and LBGs, using a robust strategy based on the identification of the Lyman break via colour excesses and photometric fits to the SEDs \citep[see][]{Barro2018}. 
The SED information has been completed using additional data from \textit{HST}/ACS, \textit{HST}/WFC3, \textit{Subaru}/Suprime-Cam, \textit{Subaru}/MOIRCS, \textit{CFHT}/WIRCam, \textit{Spitzer}/IRAC and GALEX. The sample we have built consists of 1558 candidates, separated in 124 pure LAEs with barely any continuum, 404 LBGs/LAEs and 1030 LBGs with no emission line. Within this sample, there are 92 compact groups, most of them (79) pairs of sources. Moreover, there are 10 trios, two quartets and a sextet of very close objects, fairly well distributed throughout our redshift range. For the entire sample we have calculated redshifts, SFRs, Ly$\alpha$ EWs, $X_{\mathrm{Ly}\alpha}$ and distances. Finally, we have studied the LFs, comparing them with the literature. The main conclusions are:\\

1- The characterization of the high-$z$ galaxy population ideally needs both the extremely good depth and spatial resolution only achievable with HST, and also a very good spectral resolution, as achieved with SHARDS. The main advantage of our work is the robustness of our sample. We do acknowledge that we miss a considerable amount of the faintest sources, even though SHARDS reaches a 3$\sigma$ depth of $\sim26.5$-27.0 AB mag with a seeing of $\sim0.9$ arcsec. The LFs obtained in this work are therefore a solid lower limit of the galaxy density distribution. They should be very close to the real LFs, which we estimate must be somewhere between ours and those given in previous broad band HST studies, as the latter are affected by a non negligible fraction of interlopers.\\

2- The high-$z$ candidates selection is more reliable with a large set of consecutive narrow/medium band filters than with broad band filters, as the higher spectral resolution is crucial in rejecting interlopers. We have found that, from the objects previously selected in GOODS-N within our $z$ range and magnitude limit and not affected by seeing issues, $\sim20$\% do not pass our selection criteria and are in fact lower $z$ interlopers. This is in disagreement with the previous estimation of interlopers ($\sim2$-6\%) from \citet{Vulcani2017}. We also show that this interlopers ratio slightly increases for fainter sources, from $\sim26.1$ AB mag up to the magnitudes we can reliably measure with SHARDS ($\sim27.0$ AB mag), expecting that it will be even higher for fainter magnitudes.\\

3- The $\phi^{*}$ and $M^{*}$ values found in this work for $z\sim4$ and $z\sim5$ are consistent with previous studies, since they dominate the bright region of the LFs. We are aware that we cannot build complete LFs of these populations, since we are missing the faintest sources. Therefore, our own $\alpha$ parameters cannot be derived. Nonetheless, we claim that the interloper fraction obtained using broad band surveys is sufficiently important that will incur in a small decrease in the slope of the faint end of previous LFs.\\

4- Within the objects selected in this work, some redshift inconsistencies are found between our photometric redshifts and those obtained from a colour selection criteria using broad band filters. These inconsistencies affect $\sim15$\% of the common sub-sample, highlighting the importance of good spectral information when selecting high-$z$ galaxies.\\

5- Thank to the simultaneous detection of a large number of LAEs and LBGs we have been able to obtain very good statistics of $X_{\mathrm{Ly}\alpha}$ as a function of $z$. Using a Ly$\alpha$ EW threshold of 25 \AA{}, the results are in good agreement with previous works, showing an increase of the fraction of high EW objects up to $z\sim5.5-6$. Beyond that redshift, $X_{\mathrm{Ly}\alpha}$ decreases, probably because the reionization is not completed at that epoch, thus, the increasing abundance of \ion{H}{i} scatters the Ly$\alpha$ emission line.\\

6- About 13\% of our sample appears in very close groups of two or more objects at the same redshift separated by short distances (60 kpc at maximum). The presence of tidal tail-like structures in many cases points to gravitational bounds between them. The SFR and redshift distributions of these galaxies are not different from the rest of the sample. In addition, we found 87 galaxies whose redshift is compatible  with belonging to an already reported GOODS-N overdensity at $z=4.05$ \citep{Daddi2009}. In the same way, 55 other galaxies in our sample could belong to a spectroscopically confirmed $z\sim5.198$ proto-cluster in GOODS-N \citep{Walter2012}, 44 of which have not been reported before as members of the proto-cluster. If spectroscopically confirmed, this would virtually quadruplicate the number of candidates in that overdensity, making it the richest one beyond $z=5$ up to date.

\section*{Acknowledgements}
We want to acknowledge support from the Spanish Ministry of Economy and Competitiveness (MINECO) under grants AYA2015-70498-C2-1-R, AYA2013-47742-C4-2-P and AYA2016-79724-C4-2-P. Based on observations made with the Gran Telescopio Canarias (GTC), installed in the Spanish Observatorio del Roque de los Muchachos of the Instituto de Astrofísica de Canarias, in the island of La Palma. This work is (partly) based on data obtained with the SHARDS filter set, purchased by Universidad Complutense de Madrid (UCM). SHARDS was funded by the Spanish Government through grant AYA2012-31277. PGP-G acknowledges funding by MINECO under grants AYA2012-31277, AYA2015-70815-ERC, and AYA2015-63650-P. HD acknowledges financial support from the Spanish Ministry of Economy and Competitiveness (MINECO) under the 2014 Ramón y Cajal program MINECO RYC-2014-15686. This research has made use of the NASA/IPAC Extragalactic Database (NED) which is operated by the Jet Propulsion Laboratory, California Institute of Technology, under contract with the National Aeronautics and Space Administration.\\

Special thanks to Rui Marques and Drs. David Sobral, José Alfonso López Aguerri, Jairo Méndez Abreu and José Miguel Mas Hesse for very kind and useful comments, help and discussions.

%%%%%%%%%%%%%%%%%%%%%%%%%%%%%%%%%%%%%%%%%%%%%%%%%%  

%%%%%%%%%%%%%%%%% APPENDICES %%%%%%%%%%%%%%%%%%%%%
\appendix

%%%%%%%%%%%%%%%%%%%%%%%%%%%%%%%%%%%%%%%%%%%%%%%%%%



%\bibliographystyle{mnras}


%%%%%%%%%%%%%%%%%%%%%%%%%%%%%%%%%%%%%%%%%%%%%%%%%%

% Don't change these lines
\bsp	% typesetting comment
\label{lastpage}
\end{document}